\documentclass{aa} 
\newcommand\hmmax{0}
\newcommand\bmmax{0}
\usepackage[T1]{fontenc}
\usepackage{graphicx}
\usepackage{txfonts}
\usepackage[colorlinks=true,
            linkcolor=blue,
            urlcolor=blue,
            citecolor=blue]{hyperref}
\usepackage{booktabs}
\usepackage[flushleft]{threeparttable}
\usepackage{amsmath}
\usepackage[fleqn]{amsmath}
\usepackage{nccmath}
\usepackage{microtype}
\usepackage{wrapfig}
\usepackage{upgreek}
\usepackage{textcomp}
\usepackage{subcaption}
\usepackage{bm}
\usepackage{nicematrix}
\usepackage{mathtools}

\makeatletter
\renewcommand*\aa@pageof{, page \thepage{} of \pageref*{LastPage}}
\makeatother

\def\star{_*}

\def\f1{f_{\rm I}}

\def\htwoo{H$_{2}$O\xspace}
\def\htwos{H$_{2}$S\xspace}
\def\chfour{CH$_{4}$\xspace}

\def\cotwo{CO$_{2}$\xspace}

\usepackage{xcolor}

\def\spitzer{\textit{Spitzer}\xspace}

\def\hst{\textit{HST}\xspace}
\def\jwst{\textit{JWST}\xspace}

\def\beq{\begin{equation}}
\def\eeq{\end{equation}}

\def\t2{\tau_{\rm II}}

\def\sigmas0{\Sigma_{\rm s,0}}

\def\({\left(}
\def\){\right)}
\def\<{\left<}
\def\>{\right>}

\usepackage{xparse}
\ExplSyntaxOn

\NewDocumentCommand{\DefineObj}{O{}mm}{
  \tl_new:c{g_obj_short_#2_tl}
  \tl_new:c{g_obj_long_#2_tl}
  \tl_set:cn {g_obj_short_#2_tl} {#1}
  \tl_set:cn {g_obj_long_#2_tl} {#3}
}

\NewDocumentCommand{\obj}{s m}{
  \IfBooleanTF{#1}{
    \tl_use:c {g_obj_long_#2_tl}
  }{
    \tl_if_empty:cTF {g_obj_short_#2_tl}
      { \tl_use:c {g_obj_long_#2_tl} }
      { \tl_use:c {g_obj_short_#2_tl} }
  }
}
\ExplSyntaxOff

\DefineObj[SIMP~0136]{simp}{SIMP J013656.5+093347.3}

\defcitealias{ackerman_precipitating_2001}{AM01}
\defcitealias{molliere_retrieving_2020}{M20}
\defcitealias{zhang_elemental_2023}{Z23}
\defcitealias{guillot_radiative_2010}{G10}

\begin{document} 
   \title{The JWST weather report: Unravelling the atmospheric variability of isolated worlds using principal component analysis}
    \titlerunning{Unravelling \obj{simp}'s atmospheric variability with PCA}
    \authorrunning{Schrader et al.}
   \subtitle{}

\author{Merle~A.~Schrader\inst{\ref{tcd}}\fnmsep\thanks{Corresponding author: \email{schradem@tcd.ie}}
          \and Johanna~M.~Vos
          \inst{\ref{tcd},\ref{amnh}}
          \and Evert~Nasedkin
          \inst{\ref{tcd}}
          \and Jennifer~Kestell
          \inst{\ref{tcd}}
          \and Nicolas~B.~Cowan
          \inst{\ref{mcgillPhys}, \ref{mcgillEarth}} 
          \and Roman~Akhmetshyn
          \inst{\ref{mcgillPhys}} 
          \and Samuel~Beiler
          \inst{\ref{tcd}}
          \and Beth~A.~Biller
          \inst{\ref{IoAedinburgh},\ref{CfEedinburgh}}
          \and Ben~Burningham
          \inst{\ref{Hertfordshire}}
          \and Jacqueline~Faherty
          \inst{\ref{amnh}}
          \and Eileen~C.~Gonzales
          \inst{\ref{SFSU}}
          \and Allison~M.~McCarthy
          \inst{\ref{tcd}, \ref{boston}}
          \and Caroline~V.~Morley
          \inst{\ref{utaustin}}
          \and Barry~O'Donovan
          \inst{\ref{tcd}}
          \and Cian~O'Toole
          \inst{\ref{tcd}}
          \and Genaro~Su\'arez
          \inst{\ref{amnh}}
          \and Xianyu~Tan
          \inst{\ref{tdli}}
          \and Channon~Visscher
          \inst{\ref{dordt},\ref{ssi}}
          \and Niall~Whiteford
          \inst{\ref{amnh}}
          \and Yifan~Zhou
          \inst{\ref{virginia}}
          }

\institute{
School of Physics, Trinity College Dublin, The University of Dublin, Dublin 2, Ireland
\label{tcd}
\and
Department of Astrophysics, American Museum of Natural History, Central Park West at 79th Street, New York, NY 10034, USA
\label{amnh}
\and
Department of Physics, McGill University, 3600 rue University, Montr\'eal, QC H3A 2T8, Canada
\label{mcgillPhys}
\and
Department of Earth and Planetary Sciences, McGill University, 3600 rue University, Montr\'eal, QC H3A 2T8, Canada
\label{mcgillEarth}
\and
Institute for Astronomy, University of Edinburgh, Royal Observatory, Edinburgh EH9 3HJ, United Kingdom 
\label{IoAedinburgh}
\and
Centre for Exoplanet Science, University of Edinburgh, Edinburgh, United Kingdom
\label{CfEedinburgh}
\and
Centre for Astrophysics Research, Department of Physics, Astronomy and Mathematics, University of Hertfordshire, Hatfield AL10 9AB, United Kingdom
\label{Hertfordshire}
\and
Department of Physics and Astronomy, San Francisco State University, 1600 Holloway Ave., San Francisco, CA 94132, USA
\label{SFSU}
\and
Department of Astronomy \& The Institute for Astrophysical Research, Boston University, 725 Commonwealth Avenue, Boston, MA 02215, USA
\label{boston}
\and
Department of Astronomy, University of Texas at Austin, 2515 Speedway, Austin, TX 78712, USA
\label{utaustin}
\and
Tsung-Dao Lee Institute, Shanghai Jiao Tong University, 1 Lisuo Road, Shanghai 200127, People’s Republic of China
\label{tdli}
\and
Chemistry \& Planetary Sciences, Dordt University, Sioux Center, IA 51250, USA
\label{dordt}
\and
Center for Extrasolar Planetary Systems, Space Science Institute, Boulder, CO 80301, USA
\label{ssi}
\and
Department of Astronomy, University of Virginia, 530 McCormick Road, Charlottesville, VA 22904, USA
\label{virginia}
}

   \date{Received 27 March 2026 / Accepted 10 July 2026}
\abstract{
Brown dwarf variability directly probes atmospheric dynamics beyond the Solar System, and recent \textit{James Webb Space Telescope (JWST)} time-resolved spectroscopy has opened a new window into these processes. Principal component analysis (PCA) offers a data-driven framework to identify the dominant, independent patterns of spectral variability of variable targets without relying on prior atmospheric assumptions.
\obj*{simp} (\obj{simp}) is a young, T2.5, brown dwarf at the planetary-mass boundary, making it an ideal analogue for directly imaged exoplanets. We analysed one rotation of \textit{JWST}/NIRSpec PRISM time-series spectroscopy to investigate the drivers of its variability using PCA. Prior analyses of these data revealed wavelength-dependent variability consistent with multi-layer atmospheric structure and thermal variations atop patchy clouds, motivating a complementary, model-independent approach.
Our PCA showed that the variability is intrinsically low dimensional. Two principal components are sufficient to reduce the residual spectra to the propagated noise floor, indicating that the detectable coherent spectroscopic variability is captured by the first two components. The leading principal component captures broadband variability consistent with temperature changes, while the second traces chromatic variability linked to vertical cloud structure. The existence of two dominant components implies that the spectra can be described as mixtures of three distinct atmospheric states.
We identified these three extreme spectral states from PCA projections and mapped their relative contributions as a function of rotational phase. The observed spectra are described as evolving linear combinations of these states, indicating that the variability arises from the changing visibility of spatially distinct atmospheric regions.
By projecting Sonora Diamondback forward models into the same principal component space, we found that the eigenspectra capture a large fraction of the model variance, demonstrating that the same physical processes govern much of the model grid and observed variability.
These results show that changes in temperature and cloud vertical structure account for most of the variability in \obj{simp}, and establish PCA as a computationally efficient, physically interpretable framework for analysing \textit{JWST} time-resolved spectroscopy of substellar atmospheres.
}
   \maketitle
   \nolinenumbers
\section{Introduction}
Brown dwarfs exhibit photometric and spectroscopic variability driven by atmospheric heterogeneities such as patchy condensate clouds, thermal gradients, and magnetic phenomena \citep[e.g.][]{radigan_strong_2014, Zhang_convective_2014, Hallinan2015MagnetosphericallySequence, tremblin_cloudless_2016}. Their similarity in temperature and composition to directly imaged exoplanets makes them ideal targets for high-precision, time-resolved spectrophotometry and excellent proxies for studying exoplanet atmospheres, including systems such as HR~8799~b and AF~Lep~b \citep{marois_direct_2008, Franson_AFLebb_2023}.

\spitzer and \textit{Hubble Space Telescope} (\hst)/WFC3 observations of late-L to T dwarfs have revealed rotation-modulated variability across the near- and mid-infrared. By comparing light-curve phase and maximum flux deviations as functions of wavelength these studies inferred longitudinal brightness structure (including banded patterns) and pressure-dependent phase shifts \citep[e.g.][]{buenzli_vertical_2012, apai_hst_2013,yang_extrasolarstorms_2016, apai_zonesbands_2017}. These observations motivated additional studies into the complexity of evolving exoplanetary atmospheres. \textit{JWST} now allows for continuous spectroscopic monitoring from 0.6\,$\mu$m to 5.3\,$\mu$m with NIRSpec/PRISM \citep{Jakobsen_jwst_2022}, and from 5\,$\mu$m to 14\,$\mu$m with MIRI \citep{rieke_mid-infrared_2015}, providing spectral and temporal resolution that can probe atmospheric variability over a broad range of photospheric pressures (from deep near-IR windows to higher-altitude molecular bands). \citet{biller_weather_2024}, \citet{chen_1049_2025}, and \citet{Oliveros_1049_2026} presented \jwst NIRSpec/PRISM and MIRI/LRS broad-wavelength spectroscopic light curves of the components in the benchmark binary brown dwarf WISE~J104915.57$-$531906.1AB (Luhman~16AB), revealing complex, wavelength-dependent variability. These studies point to deep patchy clouds shaping the lowest layers, higher-altitude hotspots modulating the upper atmosphere, and layer-specific variability mechanisms that appear stable across epochs.

\obj*{simp} (hereafter \obj{simp}) is a nearby \citep[$6.12\pm0.02$ pc,][]{gaia_edr3_2023}, bright, T2.5 exoplanet analogue \citep{artigau_discovery_2006} with an effective temperature of $T_{\rm eff}\sim1098\pm6$\,K \citep{zhang_coconuts_2020a}. As a member of the Carina-Near moving group, \obj{simp} is $200\pm50$~Myr old \citep{gagne_simp_2017}. Published mass estimates for \obj{simp} span a range depending on the adopted evolutionary and atmospheric modelling assumptions, from $12.7\pm1.0$ M$_{\rm Jup}$ \citep{gagne_banyan_2018} to $17.8 \pm 11.9$ M$_{\rm Jup}$ from spectral energy distribution analysis \citep{vos_patchy_2023}. \citet{vos_patchy_2023} additionally derived a surface gravity of $\log g = 4.5\pm0.4$.
\obj{simp} shows $J$-band variability with maximum deviations of up to $\sim$5\,\% \citep{artigau_photometric_2009, radigan_independent_2014} and a rapid rotation with a period of $2.41\pm0.08$~h \citep{yang_extrasolarstorms_2016}. 
Simultaneous \textit{HST} and \textit{Spitzer} monitoring reported large wavelength-dependent phase offsets (up to $\sim$180$^\circ$ from 1.1 to 1.7 and from 3.6 to 4.5\,$\mu$m), consistent with patchy cloud layers at significantly vertically separated altitudes \citep{yang_extrasolarstorms_2016}. \cite{vos_viewing_2017} measured \obj{simp}'s inclination as nearly equator-on, with $i = 80_{-12}^{+10}\,^{\circ}$. The detection of coherent radio bursts for SIMP 0136 indicate that auroral processes are present \citep{kao_aurora_2016, kao_aurora_2018}.
Near-infrared maximum flux deviations show mild wavelength dependence with reduced modulation in the 1.4\,$\mu$m \htwoo band \citep{apai_hst_2013, lew_cloudatlas_2020}. Chromatic behaviour like this is consistent with different wavelengths probing distinct pressure levels \citep{yang_extrasolarstorms_2016}. 

To help determine the physical drivers of this variability, \citet{vos_patchy_2023} applied atmospheric retrievals to a broad wavelength spectral energy distribution of \textit{SpeX} Prism \citep{burgasser_subtle_2008}, \textit{AKARI} \citep{sorahana_akari_2012}, and \textit{Spitzer}/IRS \citep{suarez_emergence_2022}. They found that models featuring vertical layers of patchy silicate and iron clouds best fit the spectra and inferred the atmospheric temperature structure and composition. Extending to \textit{JWST}, \citet{mccarthy_simp_2025} presented $0.8-11~\mu$m time-series spectroscopy using NIRSpec/PRISM and MIRI/LRS and showed that wavelength-dependent patterns arise from changes in cloud patchiness, episodic high-altitude heating, and carbon chemistry. Time‐resolved retrievals on the same dataset infer a $\sim 250$\,K inversion in the upper-atmosphere temperature \citep{nasedkin_retrieved0136_2025} which the authors plausibly linked to auroral heating \citep[see also][]{faherty_emission_2024}. They performed time-resolved retrievals, finding small ($\sim$5\,K) effective temperature changes as a function of phase. When all cloud parameters were allowed to vary, no single cloud parameter showed a strong phase trend, while the \cotwo abundance displayed phase-linked behaviour.

Building on these analyses, we applied the novel method of principal component analysis (PCA) to the same \textit{JWST}/NIRSpec PRISM time series. PCA is a data-driven framework that describes time-resolved spectra by identifying groups of co-varying wavelengths across the observations. These patterns define characteristic spectral shapes, or eigenspectra, whose linear combination re-creates the observed spectral variability \citep{cowan_alien_2009}.
The use of PCA to identify the drivers of this well-studied dataset is motivated by its unique combination of prior, independent analyses and first-of-its-kind \textit{JWST} coverage (dense time sampling over $0.6-5.3$~µm). Together these provide an ideal setup to cross-validate PCA inferences while isolating the physical mechanisms underlying the observed variability.
This technique has been applied a handful of times to brown dwarf variability data in the literature. \citet{apai_hst_2013} applied PCA to \hst/WFC3 data of \obj{simp} and 2MASS J21392676+0220226, another Carina-Near member L/T transition brown dwarf with similar $T_{\rm eff}$, over $1.1-1.7$\,$\mu$m and found that the first principal component alone accounted for $>\!99\,\%$ of the variance. This implies that the time-variable spectra in that bandpass are well described by the brightening and dimming of a single baseline spectrum (i.e., a two-component mixture where one component varies primarily in amplitude). This interpretation is inherently bandpass-limited as the WFC3/G141 range only probes the deep atmosphere \citep{mccarthy_simp_2025}. 
More recently, \citet{RomanNIRISS2025} presented PCA analysis of \obj{simp} using a single \textit{JWST} NIRISS/SOSS rotation ($0.85$--$2.83\,\mu$m), obtained $37.5$ h ($\sim$15 rotations) before the NIRSpec/PRISM visit analysed here. They found that two principal components (PCs) explain $\sim$81\,\% of the spectral variability across $0.85-2.83$\,$\mu$m and identified effective temperature and cloud vertical extent as dominant drivers of variability in the NIRISS dataset. They also recovered brightness maps that reveal a north--south asymmetry.
 
In this work we reanalysed the \textit{JWST}/NIRSpec PRISM time-series spectroscopy presented by \citet{mccarthy_simp_2025}, \citet{nasedkin_retrieved0136_2025}, and Kotten et al., (accepted, AAS), with an emphasis on identifying the drivers of \obj{simp}'s variability using PCA. Section~\ref{sec:DATA} describes the observations and data reduction.  
Section~\ref{sec:PCA} introduces the principal-component framework and applies it to the mean-subtracted spectra, establishing dimensionality and extracting the eigenspectra and their time series. In Section~\ref{sec:principal_components} we interpret the principal-component plane physically by projecting self-consistent forward models and by comparing phase-resolved projections of retrieved spectra.
Section~\ref{sec:endmembers} identifies three extreme atmospheric states that linearly combine to reproduce the data. We compare their spectral signatures to molecular cross sections and simple perturbation models, then infer longitudinal contribution maps from their temporal weights in Section~\ref{sec:mapEndmembers}. Finally, Section~\ref{sec:comparison} compares the PRISM sequence to the NIRISS/SOSS rotation obtained $37.5$\,h earlier, assessing (i) instrument-specific considerations for PCA (resolution, cadence) and (ii) the evolution of the variability between the two observations. 
The summary and conclusions are presented in Sections ~\ref{sec:summary} and ~\ref{sec:conclusions} respectively.

\begin{figure*}
\sidecaption
    \includegraphics[width=12cm]{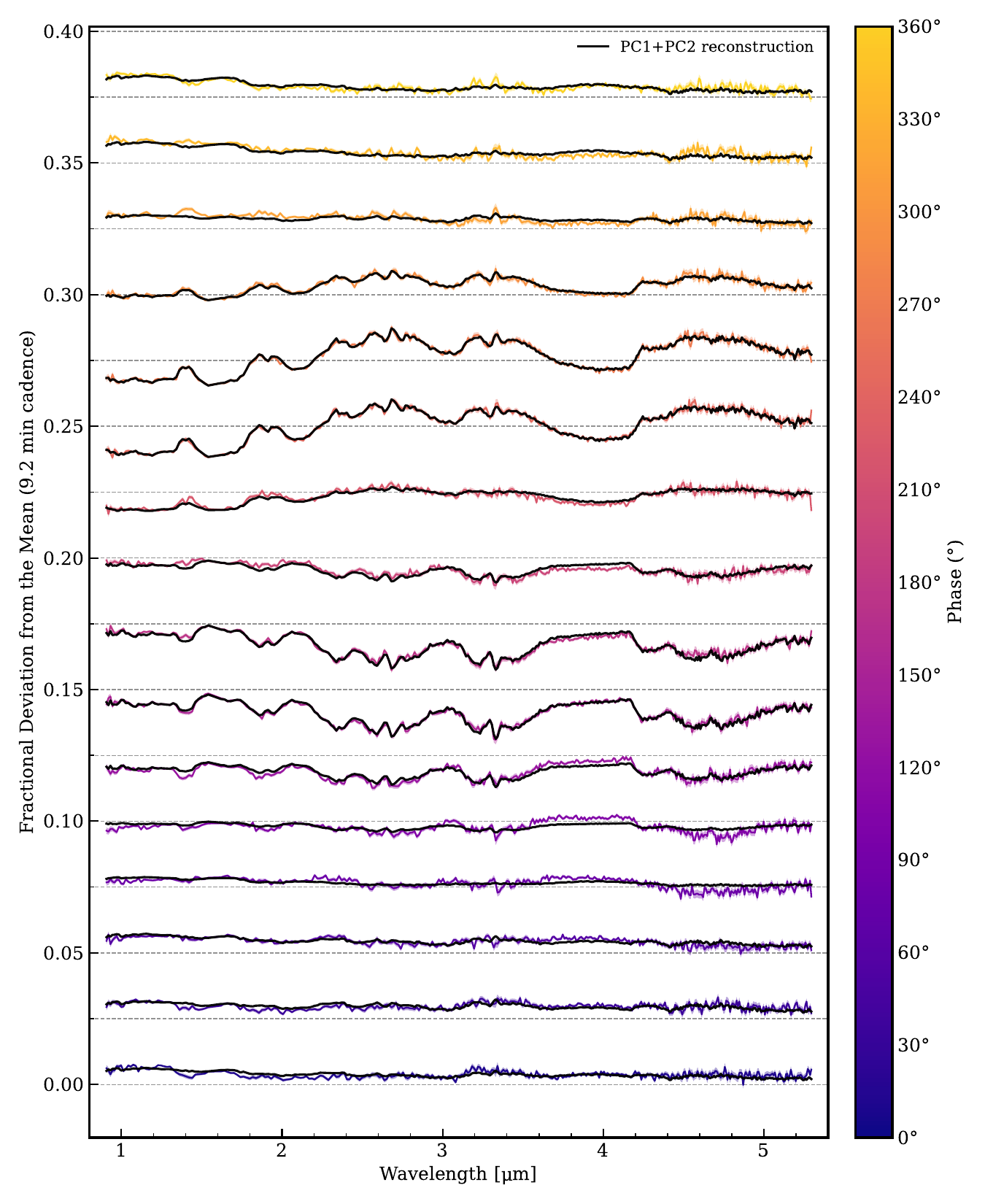}
    \caption{Fractional deviation of \obj{simp}'s spectra from the time-mean spectrum as a function of wavelength across one rotation. The black overlaid line traces the reconstructed spectra from our PCA. The spectra remain unbinned in wavelength but are binned in increments of $\sim$450 integrations along the time axis to improve visualisation.} \label{fig:DATA_spectra_vertical}
\end{figure*}

\section{Observations and data reduction}\label{sec:DATA}

We analysed time-resolved spectroscopy of \obj{simp} using the PRISM/CLEAR ($0.6-5.3\,\mu$m) configuration of NIRSpec as part of \textit{JWST} Cycle 2 program GO 3548 (PI: Vos). The observations were originally presented by \citet{mccarthy_simp_2025} and conducted in Bright Object Time Series (BOTS) mode for 1.2 rotation periods. 
The NIRSpec observations were carried out from UT 18:40:56 to 22:05:06 on 2023 July 23, resulting in a continuous time series over 2.9\,h. Observations used the SUB512 subarray with the NRSRAPID readout mode and seven groups per integration, resulting in a cadence of 1.8\,s and a total of 5726 integrations. Target acquisition was performed using the F110W filter and Wide Aperture Target Acquisition. Time-series observations with MIRI were also obtained during the same visit but are not considered in this paper. We restricted the analysis to the NIRSpec/PRISM time series because PCA requires a self-consistent set of spectra, sampling a single, continuous realisation of the atmosphere. Combining non-simultaneous observations from different instruments would require the assumption that the atmospheric state remained unchanged between epochs, which is not guaranteed for a rapidly evolving, variable object such as \obj{simp}.

The NIRSpec/PRISM time-series data were first processed with Stage~1 of the \textit{JWST} pipeline (v1.16.1)\footnote{\url{https://doi.org/10.5281/zenodo.14153298}} using the default BOTS configuration. Earlier analyses of this dataset performed two separate reductions: one optimised for flux-calibrated spectra and a second optimised for high-S/N relative light curves to avoid trade-offs between slit-loss control, background treatment, and temporal systematics \citep[][their Section 2]{mccarthy_simp_2025}.
We adopted the modified Stage~2 procedure as used by \citet{nasedkin_retrieved0136_2025}, which optimises both products in a single pass by performing a simultaneous spectral extraction (for absolute, flux-calibrated spectra) and a differential time-series extraction (for high-S/N relative light curves). Although the default BOTS Stage~2 can produce both outputs, its default configuration typically leaves the light curves noisier; the modified approach reduces these temporal residuals without degrading the absolute spectrophotometry.

To reduce low-frequency correlated noise ($1/f$) and time-dependent systematics in the time series, we processed each integration independently so that flat-fielding, wavelength
solutions, and photometric calibration were applied per time step rather than once for the whole sequence. This per-integration treatment mitigates detector drifts, pointing jitter, and pixel-to-pixel sensitivity variations that the default v1.16.1 flow does not address (it applies a single correction across the dataset). To allow for per-integration handling and enable the \texttt{nsclean} step (disabled in BOTS for v1.16.1), we changed the \texttt{EXP\_TYPE} metadata from \texttt{NRS\_BOTS} to \texttt{NRS\_FIXEDSLIT}, and then ran each integration through Stage~2 starting from the individual spectra in \texttt{rate.fits} files.

Unlike the default pipeline, which assumes one fixed spectral trace for all integrations, we fitted each integration with its own trace and aperture to capture time-variable optical distortions and shifts in the spectral position. We adopted a first-order polynomial to fit the trace as it minimises the Bayesian information criterion (BIC) among several order polynomials models, providing the simplest adequate fit.
This extraction approach improved flux calibration across the time series, ensuring accurate recovery of spectral variability. The resulting output is a time series of flux-calibrated spectra, the mean-subtracted version of which is illustrated in Fig.~\ref{fig:DATA_spectra_vertical}. We masked a persistent bad wavelength channel at $1.58~\mu$m, together with the adjacent bins on either side, because it appears as a hot-pixel feature across the time series and shows excess scatter relative to neighbouring wavelength channels that is inconsistent with the surrounding astrophysical variability. This same feature was also removed from other NIRSpec PRISM datasets in \citet{biller_weather_2024}, \citet{nasedkin_retrieved0136_2025}, \citet{mccarthy_simp_2025}, and \citet{chen_1049_2025}. We also applied conservative $3\sigma$ clipping relative to the median spectrum in time to remove isolated outlier points. We trimmed the lower and upper wavelength limits to only keep where $\mathrm{S/N}(\lambda)\ge 40$, limiting the wavelength coverage from originally $0.6-5.3~\mu\mathrm{m}$, to a range of $0.9-5.3~\mu\mathrm{m}$.

To prepare the spectra for the PCA, we interpolated them onto a common wavelength grid. PCA assumes that each wavelength bin corresponds to the same physical wavelength across all spectra. Small shifts in the trace or wavelength solution between integrations can introduce artificial variance. Interpolation was performed using the \texttt{SpectRes} routine, which conserves flux by integrating over the original pixel grid \citep{Carnall2017}.
After interpolation, we binned the spectra in time to reduce high-frequency noise and improve signal-to-noise ratio. Spectra were combined in bins of 20 integrations (corresponding to 36.6-second intervals) using an inverse-variance weighted mean, resulting in 287 spectra evenly spaced across the 2.9-hour observation. Since the STScI pipeline data-quality flags efficiently identify cosmic-ray affected pixels and other problematic samples, the weighted mean provides a statistically good estimate under approximately Gaussian noise conditions. For each time bin and wavelength channel, the uncertainty was propagated from the inverse-variance weights of the contributing spectra. The median S/N across the resulting combined and wavelength-clipped dataset is 73.0.

\section{Principal component analysis}\label{sec:PCA}

\begin{figure*}[h!]
    \centering
    \includegraphics[width=0.95\linewidth]{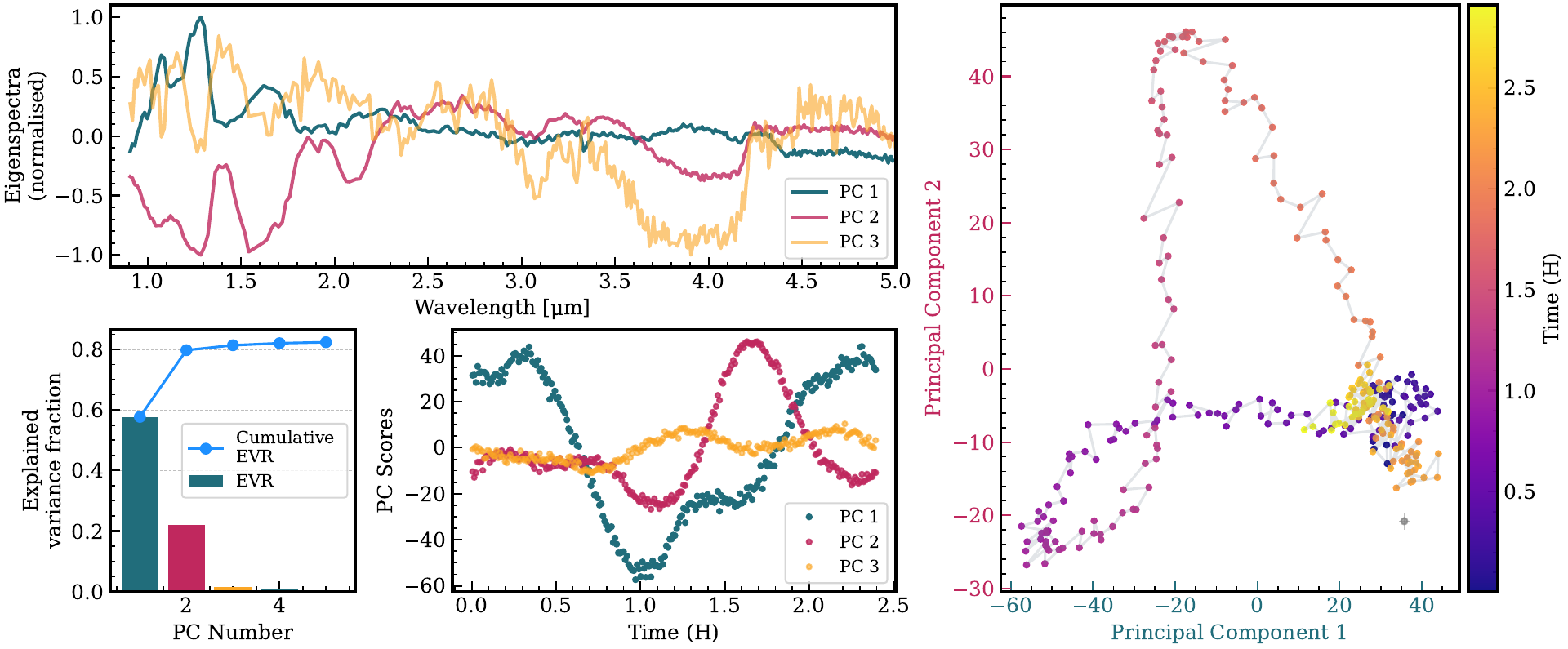}
    \caption{\textit{Top left:} The first three eigenspectra in PCA space used for the decomposition, after applying the wavelength-dependent uncertainty scaling. The first two components show broad, structured spectral covariance patterns across the PRISM range, while PC3 is less coherent and increasingly dominated by higher-frequency structure. 
    \textit{Bottom left:} The variance explained by each PC. The first two components together account for 79\,\% of the total variance in the dataset. The explained-variance fraction should not be interpreted directly as the fraction of astrophysical variability recovered because photon and detector noise also contribute to the total variance. 
    \textit{Bottom middle:} Time series of PC scores. PC1 and PC2 exhibit periodic trends consistent with rotational modulation and resemble light curves probing different atmospheric layers.
    \textit{Right:} Projection of time-resolved spectra into the plane spanned by the first two weighted PCs. Axes colours correspond to the colour-coded PCs in the three left-most plots. The PC scores and PCP axis values are amplitudes in the noise-weighted PCA space. The grey cross indicates a representative $1\sigma$ uncertainty in the projected coordinates, obtained by propagating the per-wavelength flux uncertainties through the weighted PCA projection.
    }\label{fig:PCA_analysis_subplot}
    
\end{figure*}
Principal component analysis is a statistical framework that identifies orthogonal axes of maximal variance in multivariate data \citep{Jolliffe_PCA_2016}. Conceptually, PCA relies on the idea that the observed spectra can be represented as combinations of a small number of underlying, separable patterns of variability. The observed spectra can
be interpreted as sums of contributing components. PCA assumes that the structure present in the dataset arises from a set of repeating independent modes whose linear combinations reproduce the observations \citep{Renwick_Beattie_PCA_2021}. These modes correspond directly to coherent patterns of correlated change within the data.

In the context of time-series spectroscopy, each individual spectrum can be thought of as a vector whose components are the measured fluxes in each wavelength channel. A spectrum sampled at $p$ wavelength bins therefore corresponds to a single point in $\mathbb{R}^p$, where each axis of this $p$-dimensional space represents one wavelength channel \citep{connolly_PCA_1995}. The dimensionality is thus determined directly by the number of independent spectral measurements. The direction of the resulting vector encodes the spectral shape, while its length reflects the overall brightness of the spectrum. As the object evolves over time, successive spectra trace out a collection of points in this high-dimensional spectral space.
PCA may then be understood as a rotation of the coordinate system in this spectral space so that the new axes align with the directions along which the data vary most strongly. These new axes define orthogonal modes of variability and are referred to as principal components (PCs). Each PC is itself a vector in wavelength space, with one value per wavelength channel, thus resembling a spectrum, and therefore represents a specific direction of coherent spectral change.

Importantly, PCA relies solely on the statistical information contained within the spectra themselves, without imposing prior assumptions about the atmosphere. PCA approximates each spectrum as a linear combination of a set of fixed spectral shapes (the eigenspectra of the decomposition: the PCs, $V$), the variance explained by each component ($\Sigma$) and a time series of coefficient weights (the PC scores, $U$).
We emphasise that the eigenspectra are not themselves physically observable spectra of distinct atmospheric regions. Instead, they represent particular combinations of wavelengths that are most sensitive to the different components of the atmosphere rotating in and out of view, and therefore capture patterns of variability relative to the mean spectrum. As such, they should be interpreted as describing how the spectrum changes, rather than as direct representations of individual surface or atmospheric states \citep{Cowan_unphysicalPC_2011}.

Because the weighted PCs form an orthogonal basis in the noise-whitened space, the full (mean-subtracted) spectrum can be reconstructed by summing the PCs weighted by their relevance to the variability as a whole and their PC scores as a function of time. Orthogonality in the noise-whitened space implies that each component captures variability not already described by the preceding ones: once the dominant mode (PC1) is accounted for, subsequent components describe independent patterns of change. If one were to include all components, the reconstruction would exactly reproduce the input spectrum. Using only the first few components gives an approximation that retains the dominant variability while discarding higher-order, low-variance structure.

We formalise PCA as follows: let $n$ be the number of time bins (spectra) and $p$ the number of wavelength channels. We assemble a mean-subtracted data matrix $\tilde{X} \in \mathbb{R}^{n \times p}$ by taking each spectrum and subtracting the time-averaged spectrum (the “mean spectrum”). Centring the spectra in this way ensures that the covariance matrix captures variability rather than the absolute flux level.
We performed PCA with \texttt{sklearn.decomposition.PCA}, which centres the data and computes a singular value decomposition using the \texttt{LAPACK} implementation by the method of \citet{Halko_SVD_2011}:
\begin{ceqn}
\begin{equation}
\tilde{X} = U \Sigma V^\top.
\label{eqn:SVD_1} \end{equation}
\end{ceqn}
Here, $V$ contains column vectors corresponding to orthonormal PCs in the noise-whitened space (eigenspectra): each column is a wavelength-dependent spectral shape (an axis in “spectral space”). The diagonal entries of $\Sigma \in \mathbb{R}^{n \times p}$ are the singular values, that set the variance captured by each component. The columns of $U \in \mathbb{R}^{n \times n}$ provide the PC scores, i.e., the time-dependent weights of each component. $V$ quantifies the spectral pattern of each mode; the corresponding column of $U$ quantifies how much of that pattern is present at each time; and $\Sigma$ rescales these PCs according to their variance. Truncating this sum to the leading components results in a low-dimensional, noise-reduced representation that captures the dominant variability. The spectrum at each point in time, $x_i$, may be reconstructed from the first $K<p$ PCs as a linear combination of the weighted PCs contributions, as shown below
\begin{ceqn}
\begin{equation}
\mathbf{x}_i
=
\bar{\mathbf{x}}
+
\sum_{k=1}^{K}
U_{ik}\,\Sigma_{kk}\,\mathbf{v}_k
\label{eqn:reconstruct} \end{equation}
\end{ceqn}
where $\mathbf{v}_k$ is the $k$th column of $V$, $U_{ik}$ is the PC score for component $k$ at time $i$ and $\Sigma_{kk}$ is the contribution of component k to the variability of the total dataset. The eigenspectra ($\mathbf{v_k}$ in Eqn.~\ref{eqn:reconstruct}, the columns of $V$ in Eqn.~\ref{eqn:SVD_1}) represent orthogonal modes (mutually uncorrelated patterns of change in the data) of the  spectral variability.

Because the signal-to-noise varies significantly across the PRISM wavelength range, we did not apply PCA directly to the mean-subtracted flux-space spectra. Instead, we performed a noise-weighted PCA by first whitening the spectra using the propagated per-wavelength uncertainties. For each wavelength bin $j$, we estimated a characteristic uncertainty scale, $s_j$, from the median propagated uncertainty across the time series, and constructed a noise-whitened matrix
\begin{ceqn}
\begin{equation}
\tilde{X}'_{ij}
=
\frac{\tilde{X}_{ij}}{s_j}.
\label{eqn:weighted_matrix}
\end{equation}
\end{ceqn}
The decomposition was then performed on $\tilde{X}'$ rather than directly on $\tilde{X}$. This weighting downplays wavelength regions dominated by larger formal uncertainties and ensures that the decomposition is driven primarily by coherent spectroscopic variability rather than by high-variance noise. 
The resulting PCs are orthogonal and statistically uncorrelated within the noise-whitened space in which the decomposition is performed. In this weighted space, the reconstruction is
\begin{ceqn}
\begin{equation}
\tilde{\mathbf{x}}'_{i,K}
=
\sum_{k=1}^{K}
U_{ik}\,\Sigma_{kk}\,\mathbf{v}_k,
\end{equation}
\end{ceqn}
and the corresponding reconstructed spectrum in the original flux units is obtained by reversing the whitening,
\begin{ceqn}
\begin{equation}
\mathbf{x}_{i,K}
=
\bar{\mathbf{x}}
+
\mathbf{s}\odot
\sum_{k=1}^{K}
U_{ik}\,\Sigma_{kk}\,\mathbf{v}_k,
\label{eqn:weighted_reconstruct}
\end{equation}
\end{ceqn}
where $\mathbf{s}$ is the vector of wavelength-dependent uncertainty scales and $\odot$ denotes element-wise multiplication. When transformed back into flux space for interpretation, the eigenspectra are no longer guaranteed to remain exactly orthogonal under the standard flux-space inner product because the wavelength-dependent weighting alters the metric of the space. In practice, because the whitening consists only of diagonal per-wavelength scaling, the resulting PCs remain close to orthogonal and continue to isolate approximately independent modes of variability. Throughout this work we therefore retain the standard PCA terminology and interpret the weighted PCs as independent variability modes to first order.

We applied PCA to a single full rotation (the initial $2.4$h) to prevent the PCs from being biased by duplicated observations across the same phase ($0.5\,$h or 72$^\circ$ overlap). To assess the nature of the variability captured by the PCs, we examined the first three eigenspectra in the top panel of Fig.~\ref{fig:PCA_analysis_subplot}. These eigenspectra are shown in the PCA space defined above, after the wavelength-dependent uncertainty scaling has been applied. They therefore show the spectral patterns used by the decomposition to identify coherent variability, rather than flux-space deviations with physical units. The first two components show broad, structured wavelength-dependent features, while PC3 is less coherent and is increasingly dominated by higher-frequency structure. We note that PCA does not require different components to occupy different wavelength regions. Instead, it identifies independent directions of covariance across the full spectrum; two PCs can therefore both show large structure in similar wavelength ranges while still describing different variability patterns through their full spectral shapes and time-dependent scores. This underlines PCA's strength in isolating independent drivers of variability that may dominate the same wavelength space.

In the bottom left of Fig.~\ref{fig:PCA_analysis_subplot} we show the noise-weighted variance in the dataset explained by the PCs. The bar heights give the fraction of total variance explained by each successive PC ($\Sigma^2$), while the overplotted curve shows the cumulative variance explained. The first two components together account for $79\,\%$ of the total variance in the dataset. Because photon and detector noise contribute to the total variance, the explained-variance fraction alone does not directly measure the fraction of astrophysical variability recovered. We therefore assess the dimensionality of the variability by comparing the PCA reconstructions with the propagated noise floor. To quantify the contribution of each component to the observed variability, we calculated the root-mean-square (RMS) amplitude of the reconstructed spectral variations associated with each PC across the time series. The first three weighted PCs have RMS amplitudes of $0.43\%$, $0.29\%$, and $0.07\%$, respectively, demonstrating that the amplitude of the variability captured by successive PCs drops sharply beyond the first two components. After removal of the first two weighted PCs, the residual spectra have an RMS scatter of $0.36\%$, consistent with the propagated noise floor of $0.37\%$, with no measurable excess RMS above the noise and a reduced $\chi^2$ of $0.97$. The median temporal lag-1 correlation of the noise-normalised residuals also decreases from $0.74$ before PC subtraction to $0.08$ after removing two PCs. We therefore found no evidence for additional coherent spectroscopic variability beyond the first two weighted PCs, motivating our choice to define the principal component plane (PCP) by PC1 and PC2. This indicates that the spectroscopic variability is intrinsically low-dimensional and can be captured accurately with just two axes.

The PC scores, shown in the bottom-middle panel of Fig.~\ref{fig:PCA_analysis_subplot}, are the time-dependent amplitudes of each component ($U_{ik}$ in Eqn.~\ref{eqn:reconstruct} and the $i$th columns of $U$ in Eqn.~\ref{eqn:SVD_1} at the $i$th point in time). The PC scores trace how the contribution of each spectral pattern evolves with time. Because the PCA is performed in noise-whitened space, the PC scores are amplitudes in the weighted PCA coordinates. 
%The corresponding reconstructed PC contributions are converted back into flux units by multiplying by the wavelength-dependent uncertainty scale, $s_j$, as in Eqn.~\ref{eqn:weighted_reconstruct}.
The near-periodic structure in PC1 and PC2 suggests rotational modulation. The dominant periodicity in both the first and second PCs' scores is consistent with the measured rotation period of \obj{simp}. %\citep[$2.41\pm0.08$~h;][]{yang_extrasolarstorms_2016}.

The principal-component plane (PCP) is the two-dimensional space spanned by the contribution of the first two PCs, PC1 and PC2. Projecting a spectrum into this plane means describing it by the pair of coordinates that quantify how much each component is needed to reconstruct the observed spectrum at a point in time. We note that a negative PC score implies that the spectrum is displaced in the direction opposite to that defined by the corresponding eigenspectrum. We projected all binned spectra into the PCP (not just the first rotation that the PCA was trained on). Each point in the right panel of Fig.~\ref{fig:PCA_analysis_subplot} is one spectrum observed at one time step reconstructed using only the first two PCs in the weighting of the axes coordinates. In this view, distance and direction across the PCP encode how the spectrum changes relative to the noise-weighted PCA basis. 
We estimated uncertainties on the PC scores by propagating the flux uncertainties through the weighted PCA projection. Prior to applying PCA, each mean-subtracted spectrum was divided by a wavelength-dependent uncertainty scale, $s_j$, estimated from the median propagated uncertainty in each wavelength bin. The projection onto a weighted PC is therefore a linear operation in this noise-whitened space. If the flux uncertainties in different wavelength bins are treated as independent, the variance of the score along component $k$ is
\begin{ceqn}
\begin{equation}
\sigma_{\mathrm{PC},k}^{2}
=
\sum_{j=1}^{p}
\left(
v_{kj}\frac{\sigma_j}{s_j}
\right)^2,
\end{equation}
\end{ceqn}
where $v_{kj}$ is the $j$th element of the $k$th weighted PCA basis vector, $\sigma_j$ is the $1\sigma$ flux uncertainty in wavelength bin $j$ for a given spectrum, and $s_j$ is the wavelength-dependent scale used to whiten the data. We evaluated this for PC1 and PC2 using the propagated uncertainty spectrum, which gives representative uncertainties in the weighted PCP, illustrated by the grey uncertainty cross shown in Fig.~\ref{fig:PCA_analysis_subplot}. The observed width of the trajectory in PC space is somewhat larger than this estimate, likely due to a combination of residual correlated noise (including low-level residual $1/f$ structure), imperfect whitening of wavelength-dependent systematics, and additional low-amplitude variability not fully captured by the first two PCs.

\section{Principal components as drivers of atmospheric variability}\label{sec:principal_components}
The eigenspectra and their time-dependent PC scores isolate a small number of dominant variability patterns in the spectra. For \obj{simp}, the first two weighted PCs contain the detectable coherent spectroscopic variability: after subtracting these two components, the residual spectra are consistent with the propagated noise floor.

\begin{figure*}
\centering
\begin{minipage}[t]{.33\textwidth}
\centering
    \includegraphics[width=1.0\linewidth]{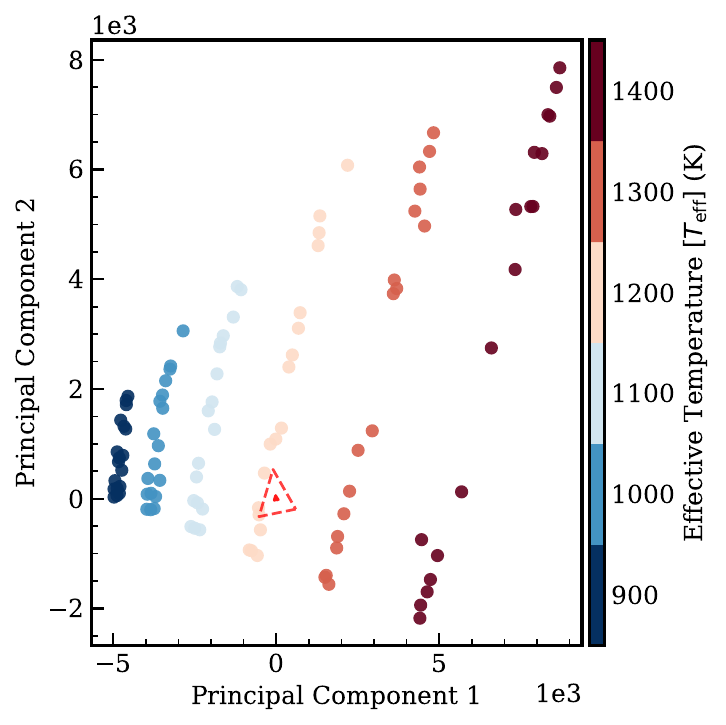}
\end{minipage}\hfill
\begin{minipage}[t]{.33\textwidth}
\centering
    \includegraphics[width=1.0\linewidth]{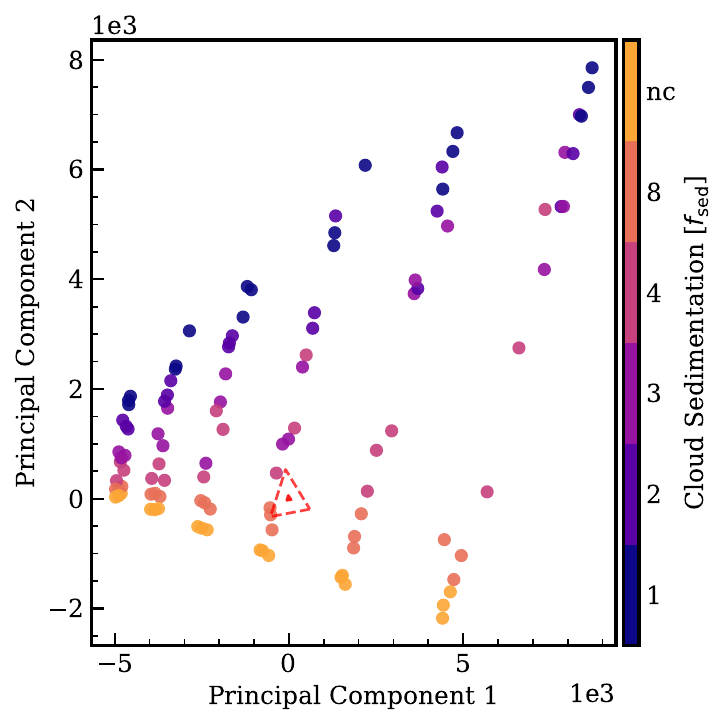}
\end{minipage}\hfill
\begin{minipage}[t]{.33\textwidth}
\centering
    \includegraphics[width=1.0\linewidth]{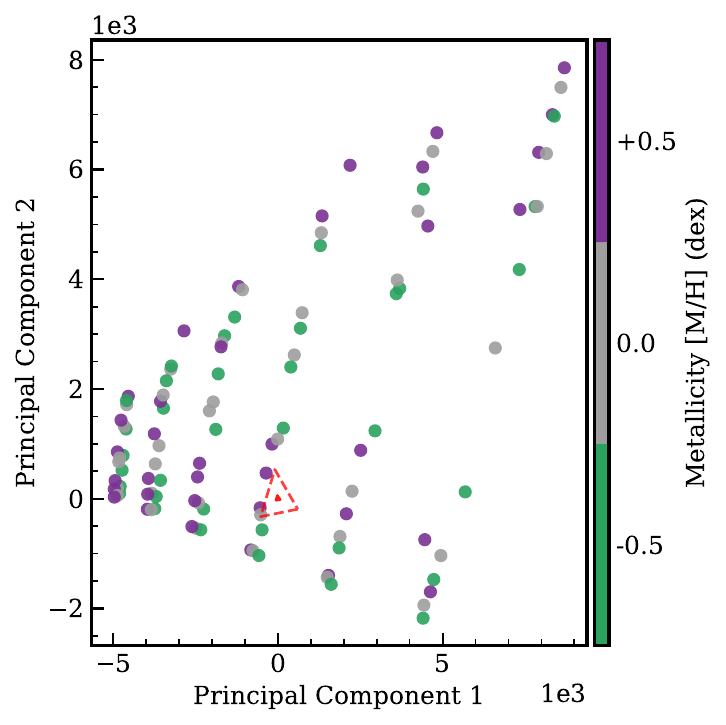}
\end{minipage}
\caption{Sonora Diamondback models projected into the PCP. Model fluxes were scaled by matching the $T_{\rm eff}=1200$\,K, $\log{g}=4~{\rm [cm~s^{-2}]}$, $f_{\rm sed}=3$, ${\rm [M/H]}=0.5$ model to the median observed spectrum. The same mean subtraction and noise-whitening used for the observed spectra were applied before projecting the models into the PCP spanned by \obj{simp}'s first two PCs. \textit{Left:} $T_{\rm eff}$ varies primarily along PC1. \textit{Middle:} $f_{\rm sed}$, spanning from 1 (low cloud sedimentation: thicker clouds) to nc (`no clouds'), varies primarily along PC2, with larger $f_{\rm sed}$ corresponding to thinner and more settled clouds. \textit{Right:} metallicity shows no obvious trend. The dashed-line red triangle marks the shrink-wrapped simplex enclosing the data, enlarged by ten times. This illustrates how coarse the model grid is in comparison to the data locus. The red, solid-line triangle is the original triangle simplex.}
\label{fig:MODELS_inPCP}
\end{figure*}

\citet{mccarthy_simp_2025} examined the same time-series spectroscopy presented in this paper and clustered the multi-wavelength light curves into groups of similar morphology. They identified 9 distinct light curve shapes and connected the clusters to specific pressures at specific altitudes in the atmosphere using contribution functions from Sonora Diamondback models \citep{morley_diamondback_2024}. Their clustering of spectroscopic light-curve morphologies by wavelength therefore maps onto distinct pressure ranges, interpreted as arising from a combination of deeper patchy clouds, higher-altitude hot spots, and chemical variations in the atmosphere.

Applying this context to the PCA results, we interpret the first two PCs as dominant full-spectrum variability patterns whose time dependence can be compared to the clustered light curves of \citet{mccarthy_simp_2025}. 
The PC-score morphologies show qualitative similarities to the clustered spectroscopic light curves reported by \citet{mccarthy_simp_2025}: PC1 exhibits a broad modulation with two local minima over the rotation, while PC2 shows a lower-amplitude, phase-shifted structure with a local minimum followed by a local maximum. The detailed relative amplitudes and phase structure, however, do not match individual clusters one-to-one. This is expected because the noise weighting changes the relative contribution of different wavelength regions to the covariance structure, increasing the influence of higher-S/N channels while downweighting noisier parts of the PRISM range. In our separate analysis using unweighted PCA, where all wavelength channels contribute equally, the PC-score time series more closely resemble the clustered light curves of \citet{mccarthy_simp_2025}. In the weighted analysis used here, the large-amplitude, smoothly varying PC1 score nevertheless shares the broad morphology of their deeper clusters, which they associate with pressures of order ${\sim10}$~bar, while PC2 shows a phase-shifted morphology qualitatively consistent with variability from higher atmospheric layers. This correspondence suggests that the weighted PCA separates broadly similar physical drivers to those identified by the light-curve clustering analysis: PC1 primarily traces deeper, continuum-dominated variability, while PC2 emphasises higher-altitude and molecular-band-sensitive variability.

This also reframes the phase lags reported in earlier multi-band variability studies. Simultaneous \hst/WFC3 and \spitzer/IRAC monitoring of brown dwarfs revealed large phase offsets between near- and mid-infrared light curves, including the $\sim180^\circ$ lag reported for \obj{simp} by \citet{yang_extrasolarstorms_2016} and similar behaviour in cooler T dwarfs \citep{buenzli_vertical_2012}. In our PCA interpretation, these wavelength-dependent lags are not necessarily evidence for one atmospheric pattern observed at different wavelengths with a simple phase shift. Rather, each broad-band light curve is a different projection of the same low-dimensional variability: a filter that is more sensitive to the PC1-like component will follow one morphology, while a filter that is more sensitive to the PC2-like component will follow another. Intermediate spectral regions naturally produce intermediate light-curve shapes, consistent with the clustering results of \citet{mccarthy_simp_2025}. Thus, the historical `phase lag' picture can be understood as the observational consequence of mixing multiple physically distinct variability modes, likely associated with different pressure-sensitive regions of the atmosphere.

The spiral-looping track in the bottom-left of the PC1--PC2 plane in the right plot in Fig.~\ref{fig:PCA_analysis_subplot} marks an interesting phase pattern. As the deep layers dim and rotate out of view (PC1 decreases), the higher layers reach their minimum contribution slightly earlier or later (PC2's local minimum is at a different phase). That small timing offset appears as a lag between the PC1 and PC2 score time series. In the PC plane, the lag appears as a smooth loop, supporting the hypothesis that the signals from different altitudes are varying in phase.

\subsection{Model projections into the PCP}\label{sec:modelsInPCP}

To interpret the physical processes reflected in the PCP, we projected a grid of the Sonora Diamondback self-consistent atmospheric models \citep{morley_diamondback_2024} into the PCP, broadly following the approach presented by \citet{RomanNIRISS2025}. 
%same space defined by the first two PCs derived from the \obj{simp} time series spectra.
The grid provides one-dimensional radiative-convective equilibrium models of brown dwarf and giant exoplanet atmospheres, spanning a range of effective temperatures, surface gravities, metallicities, and cloud sedimentation efficiencies, assuming chemical equilibrium and solar C/O ratios. 
Cloud vertical extent and optical depth follow the \citet{ackerman_marley_2001} description through the sedimentation parameter $f_{\mathrm{sed}}$, which regulates the balance between upward turbulent mixing and gravitational settling. Low $f_{\mathrm{sed}}$ indicates lower sedimentation efficiency of the clouds, resulting in thicker, vertically extended clouds; high $f_{\mathrm{sed}}$ yields optically thinner clouds \citep{ackerman_marley_2001, morley_diamondback_2024}.

For consistency with recent analyses of SIMP~0136, we fixed $\log g$ to the value adopted in \citet{RomanNIRISS2025}. The choice of surface gravity only has a minor effect on the model projections in principal-component space, and we therefore adopted this value for consistency rather than as a fitted parameter.
Each model spectrum was scaled to the observed flux level using a single multiplicative factor derived from the best-fitting time-averaged spectrum of \citet{RomanNIRISS2025} ($T_{\rm eff}=1200\,$K, $\log{g}=4~{\rm [cm~s^{-2}]}$, $f_{\rm sed}=3$, ${\rm [M/H]}=0.5$). We determined a scalar that minimises the squared difference between this reference model and the observed mean spectrum, and applied this same factor uniformly to all models in the grid. 
Across the model grid, we varied effective temperature ($T_{\rm eff}$), cloud sedimentation efficiency ($f_{\rm sed}$), and metallicity ([M/H]). We restricted the projected models to $900<{T_{\mathrm{eff}}\text{(K)}}<1400$. This range includes the measured $\sim1100$\,K of \obj{simp} \citep{vos_patchy_2023}, brackets the model we used to calculate the scaling factor and encompasses the L/T-transition regime.

The scaled model spectra were then mean-subtracted using the observational mean spectrum, divided by the same wavelength-dependent uncertainty scale used to whiten the data, and projected into the PCP using the PCA basis derived from the observations, as described in Appendix~\ref{sec:app_modelproj}. 
We subsequently calculated how well the PCA basis derived from \obj{simp}'s variability reconstructs the variance across the model grid. The first PC alone reconstructs $35.9\,\%$ of the model variance, while the first two PCs reconstruct $79.9\,\%$. Adding PC3 increases this only marginally to $80.2\,\%$. We therefore interpret that the first two PCs are capturing the model variations that are most strongly sampled by the observed time variability. 
Despite the simplifying assumptions of the grid models (such as being 1D plane-parallel, assuming radiative-convective equilibrium and equilibrium chemistry), the fraction of model variance reconstructed by the first two data-driven PCs is high. %, inspiring confidence in our physical interpretation of the PCP. 
This agreement suggests that the dominant axes of variability identified by PCA correspond to the same physical processes that differentiate models across the Diamondback grid. In particular, changes in $T_{\rm eff}$ and $f_{\rm sed}$, which are the parameters that change the model spectra, map onto the same dominant variability modes recovered from the data. Although the grid itself is discrete and relatively coarse, our results show that the physics driving the differences between neighbouring models is closely related to the processes governing atmospheric variability in \obj{simp}. The forward models can therefore not only describe static atmospheric states but also contain the underlying physics of variability.

Examining the distribution of the model grid within the PCP shows some clear trends, presented in Fig.~\ref{fig:MODELS_inPCP}. Progression along the PC1 axis increases $T_{\rm eff}$, while a progression along PC2 increases $f_{\mathrm{sed}}$. This suggests that the first PC is strongly correlated with brightness-driven variability ($T_{\rm eff}$), while the second captures changes in cloud vertical structure. No discernible trend is observed with metallicity, which is expected given that it is a bulk atmospheric property and is not expected to vary spatially across a single object. We verified that the same qualitative trends are recovered with an unweighted PCA, although the noise-weighted decomposition yields a cleaner separation between the temperature-like and cloud-sedimentation directions.

We interpret the variation in $T_{\rm eff}$ with PC1 as changes in the thermal structure as opposed to changes in the internal energy of the object. For a convective brown dwarf, the interior entropy and deep adiabat are not expected to vary significantly on rotational timescales \citep{saumon_evolution_2008}. Instead, variations in the observed flux arise from spatial and thermal heterogeneities in the atmosphere including evolving atmospheric structures, and dynamical processes such as convection and large-scale circulation. These processes can modulate the radiative flux emerging from different regions in the atmosphere, producing changes in the disk-integrated brightness as the object rotates. Heterogeneous opacity sources, including but not limited to the extent of patchy clouds, further shape the depth and efficiency of emission across wavelengths \citep{marley_review_2015, tan_atmospheric_2021}. In this context, motion along the PC1 axis reflects changes in the observed flux distribution driven by atmospheric structure and dynamics, which we interpreted as changes in the atmosphere between hotter and cooler thermal profiles and refer to as temperature.

Notably, the self-consistent grid models lack the parameter resolution to resolve the subtle differences between our observed spectra that likely stem from small atmospheric patches. The red triangle in Fig.~\ref{fig:MODELS_inPCP}, which represents a 10x enlargement of the smallest area triangle that can be fit around the data locus, remains much smaller than the spread of model projections in the PCP. This highlights PCA's ability to isolate qualitative, physically meaningful trends in the spectra that lie below the current resolution of self-consistent grids.
Although the coarseness of the grid model projection does not allow us to directly trace how these parameters change with phase, the trends we observed allow for a qualitative interpretation of the triangular trajectory traced by the observed spectra in the PCP. 
The rotation of \obj{simp} produces a cyclic modulation in PC space, first along an axis of consistently low $f_{\mathrm{sed}}$ and decreasing temperature only to then move along an axis of consistent temperature but decreasing $f_{\mathrm{sed}}$. Finally, the $T_{\rm eff}$ and $f_{\mathrm{sed}}$ increase again to close the loop where the rotation began. 

The trends identified in the model projections provide a physical interpretation of the PCP, where PC1 primarily traces temperature-like variability and PC2 captures cloud vertical extent. The model-projected $T_{\rm eff}$ and $f_{\rm sed}$ directions are nearly aligned with PC1 and PC2, respectively, but they are not expected to align perfectly. PCA identifies orthogonal directions of maximal covariance in the mean-subtracted, noise-whitened spectra, whereas the physical parameters in the model grid are not statistically independent. In particular, $T_{\rm eff}$ and cloud properties are intrinsically coupled: changes in cloud opacity modify the atmospheric temperature structure and contribution functions, leading to correlated spectral responses \citep[e.g.][]{ackerman_marley_2001, saumon_evolution_2008, marley_review_2015}. As a result, variations across the model grid do not map onto perfectly orthogonal directions in spectral space. These correlated changes in the model grids produce the small misalignment between the PCA axes and the $(T_{\rm eff}, f_{\rm sed})$ directions, while still providing a physical interpretation of the dominant modes of variability.
Our results are consistent with those of \citet{RomanNIRISS2025}, who present a NIRISS/SOSS time series taken $37.5$\,h earlier. Their PCs show similar trends in the PCP, with physically interpretable axes corresponding to $T_{\rm eff}$ and $f_{\rm sed}$. Because their analysis used an unweighted PCA, the explained-variance fractions are not directly comparable to those from our weighted PRISM analysis. We therefore focus on the qualitative agreement in the physical interpretation of the PC axes, and present an in-depth comparison of PCA applied to both datasets in Section~\ref{sec:comparison}.

\subsection{Retrieved model projections in the PCP}

To compare the PC-projected spectra to quantitatively inferred atmospheric properties, we projected the time-resolved best-fit retrieved spectra from \citet{nasedkin_retrieved0136_2025} into the same PCP defined by our PRISM observations. In their study, 24 phase-resolved spectra, spanning the same PRISM time series analysed in this work, were fit with the \texttt{petitRADTRANS} retrieval framework \citep{molliere_petitradtrans_2019,blain_prt_2024}. These retrievals fit the combined NIRSpec/PRISM and MIRI/LRS spectra using a forward model that used temperature profile parameterisation of \cite{zhang_elemental_2023}, patchy silicate clouds with an iron deck, and freely retrieved chemical abundances. The CH$_{4}$ and CO abundances were allowed to vary as a function of pressure. The models were calculated at a spectral resolving power of 400 using the correlated-k method, before being convolved by the instrumental line-spread function and binned to the observed spectra. For each retrieved spectrum, we (i) convolved to the NIRSpec/PRISM resolving power using the published instrumental  line spread functions, (ii) flux-conserve re-binned onto the exact wavelength grid used for PCA, and (iii) applied the same mean-subtraction, masking and noise-whitening as the data before projecting with the observed eigenspectra. This ensures a self-consistent comparison between the observations and the retrieved models. The NIRSpec/PRISM and MIRI/LRS spectral resolution functions were taken from CRDS and applied via a Gaussian kernel to binning, see \citet{nasedkin_retrieved0136_2025}'s Sect.~3 and Eq.~(1) for the likelihood on the binned spectra.

\begin{figure*}[h]
   \centering
   \includegraphics[width=\textwidth]{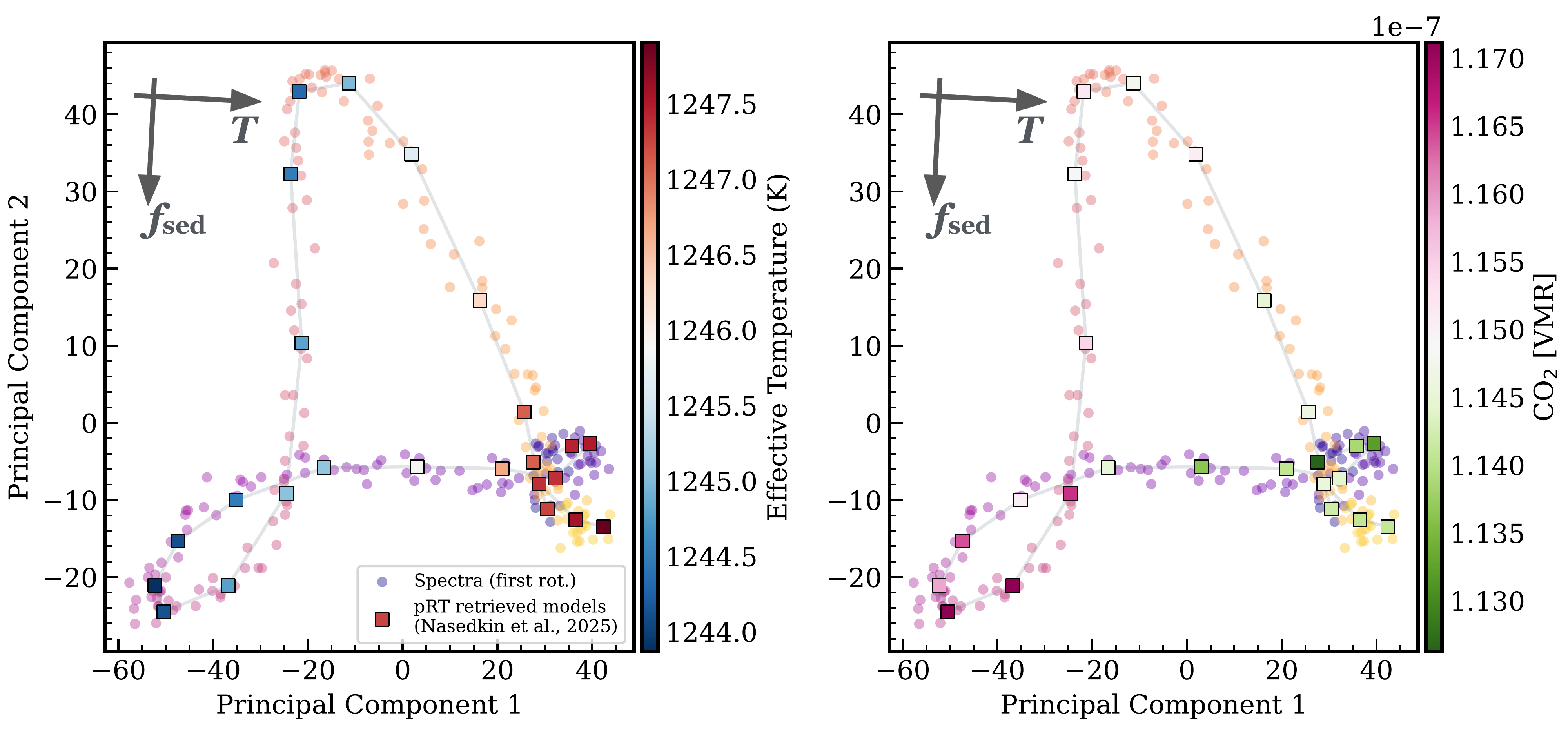}
   \caption{Parameters retrieved by \citet{nasedkin_retrieved0136_2025} projected into the PCP, coloured by \textit{Left:} Effective temperature (high $T_{\rm eff}$ = red); \textit{Right:} \cotwo abundance (high \cotwo abundance = purple); %\textbf{Right:} \htwos (high \htwos abundance = brown).
   Effective temperature varies predominantly along PC1. $\mathrm{CO_2}$ exhibits an inverse trend to $T_{\rm eff}$ that is significantly more pronounced where clouds are thinner (high $f_{\rm sed}$).
   }
   \label{fig:PCA_retr_colourcode}
\end{figure*} 

The retrieved models do not perfectly reproduce the observed spectra, with specific wavelengths that are systematically poorly fit across all 24 retrievals \citep[][see Sect.~4.3, Sect.~5.1, and Fig.~C.4]{nasedkin_retrieved0136_2025}. When PCA is applied, this produces a constant offset in PC space, reflecting a difference in the mean value of the retrieved spectra and the data we use to train the PCA. In the context of this paper, our goal is to assess whether the evolution of the spectra through the PCP is reproduced by the retrieved models. We therefore subtracted this difference in the PC projection of the mean retrieved spectrum from the PC components of the retrieved spectra.
After this translation, the retrieved sequence closely follows the observed triangle in PCP as illustrated in Fig.~\ref{fig:PCA_retr_colourcode}, showing that the variability captured in the first two PCs is characterised by the retrievals. After projecting the phase-resolved retrieved models into the PCP, we colour-coded each model point by the retrieved effective temperature $T_{\rm eff}$ and \cotwo\ abundance, respectively. The same process is applied to the \htwos, CO, and \chfour abundances (Fig. \ref{fig:appendix_retr_PCA}). We focus our discussion primarily on $T_{\rm eff}$, \cotwo\ and \htwos\ as these parameters varied in the most significant, least statistically random way \citep{nasedkin_retrieved0136_2025}. 

The left panel of Fig.~\ref{fig:PCA_retr_colourcode} shows that the retrieved effective temperature varies along the axis we identified as the Sonora forward-model temperature axis, which closely aligns with PC1. This alignment is expected. In L/T-transition atmospheres, deeper heating primarily modulates the near-IR continuum windows and therefore projects most strongly onto PC1 (see Sect.~\ref{sec:modelsInPCP}; see also \citealt{ackerman_marley_2001, mccarthy_simp_2025}). The consistency between the retrievals and forward models demonstrates that PC1, which reflects the dominant axis of variability identified by PCA, reflects real changes in the emitted flux.

The retrieved \cotwo\ abundance (right panel, Fig.~\ref{fig:PCA_retr_colourcode}) shows a weaker, generally anti-correlated trend with PC1 (higher \cotwo\ abundance at lower PC1). The trend is clearest during phases that the Sonora Diamondback grids indicate have optically thinner clouds (high $f_{\rm sed}$, i.e.\ lower along PC2), where abundance-driven spectral differences are more readily visible. When clouds are more vertically extended (lower $f_{\rm sed}$), this trend persists but is significantly muted.

One scenario that could give rise to the observed phase-dependent \cotwo\ variability is a change in the effective photospheric depth driven by clouds. Variations in cloud opacity shift the pressure level from which we observe radiation, meaning that different phases probe different layers of the atmosphere \citep[e.g.][]{apai_hst_2013}. Because temperature and chemical abundances are expected to vary with depth, this can produce an apparent change in \cotwo\ even if the bulk abundance is unchanged \citep{Lee_h2s_2024}.
Alternatively, the variability could reflect changes in the local temperature-pressure structure. Atmospheric dynamics and radiative feedback from clouds can modify the thermal profile \citep{tan_atmospheric_2021}, and the partitioning of carbon between CO, \chfour, and \cotwo\ is highly temperature dependent. In general, higher temperatures and pressures favour CO and \cotwo\ while lower temperatures and pressure favour \chfour\ \citep{zahnle_marley_2014}. Changes in the thermal structure at the pressures probed by the observations could therefore shift this balance and produce phase-dependent \cotwo\ variability.
Finally, it is also possible that the \cotwo\ variability is driven by changes in vertical mixing. In this case, abundances become quenched at the pressure level where chemical and mixing timescales are equal. At deeper levels, chemical timescales dominate, while higher in the atmosphere, mixing can transport material faster than the chemistry can reach equilibrium. The strength of this mixing is commonly parametrised by the eddy diffusion coefficient, $K_{zz}$, which quantifies the efficiency of vertical transport in the atmosphere. Changes in $K_{zz}$ shift the pressure level at which species are quenched, thereby altering the abundances observed at photospheric levels. The \cotwo\ abundance at observable pressures is influenced both by where it quenches and the CO reservoir, which quenches at deeper levels \citep{beiler_twoMols_2024, wogan_update_2025}. Variations in $K_{zz}$ can therefore shift the pressure, and therefore abundance, at which \cotwo\ is quenched. Self-consistent models show that increasing $K_{zz}$ can lead to measurable changes in \cotwo\ abundance at observable pressures (e.g.\ Fig.~7 of \citealt{mukherjee_probing_2022}), illustrating how variations in vertical mixing alone may contribute to the phase-dependent behaviour.

These scenarios are not necessarily independent. Cloud opacity, thermal structure, and chemistry are intrinsically coupled: clouds influence the temperature profile, the temperature profile sets the chemical balance, and vertical mixing determines how that chemistry is transported and preserved. The observed \cotwo\ behaviour may be driven by a combination of these processes and the present data do not allow us to distinguish cleanly between them.
If the variability is driven by carbon chemistry or changes in cloud opacity, correlated changes in CO and \chfour\ are expected. However, we did not detect a clear and coherent phase trend in the retrieved CO or \chfour\ abundances (see Section~\ref{sec:app_retrievals}). This does not rule out a chemical origin, but suggests either that the effect is subtle, or that degeneracies in the retrieval limit our sensitivity to variability in these species.

The Sonora Diamondback forward-model projections indicate cloud-linked variability via $f_{\rm sed}$ (Section ~\ref{sec:modelsInPCP}). 
In the retrievals of \citet{nasedkin_retrieved0136_2025}, however, explicit tests show that (i) fixing the temperature profile cannot reproduce the observed variability, whereas (ii) fixing cloud parameters still results in comparably good fits. Moreover, no statistically significant phase trend is recovered in individual cloud parameters.
This apparent discrepancy may reflect differences in parameterisation. The Diamondback models compress cloud physics into a single parameter ($f_{\rm sed}$), producing a clear, interpretable trend. The retrievals instead allow multiple cloud parameters (coverage, base pressure, particle size, mass fraction), which introduces degeneracies that can absorb variability in different ways. As a result, cloud-driven variability may not manifest as a clear trend in any single retrieved parameter, even if clouds play a physical role.

In summary, the alignment between PC1 and $T_{\rm eff}$, the phase-dependent modulation of \cotwo, and the cloud-dependent suppression of its variability are consistent with a picture in which temperature, cloud structure, and chemistry may all contribute to the variability. PCA provides an efficient framework in which these effects can be explored while remaining agnostic about the specific physical mechanisms driving the variability.

\section{Spectral endmembers as atmospheric extreme states}\label{sec:endmembers}

To interpret the structure of the variability in the PCP, we first note that the two-dimensional space implies that each of the reconstructed spectra in the PCP can be described as a mixture of at least three distinct spectral surface types, which we refer to as spectral endmembers. These endmembers represent extreme, hypothetical atmospheric states rotating in and out of view. More generally, in an $N$-dimensional PC space the minimal simplex shrink wrap has $N+1$ vertices that enclose all projected spectra \citep{Cowan_unphysicalPC_2011, Cowan_2013}. In our case of two dimensions, this simplex is a triangle. The three vertices represent distinct spectral states with extreme atmospheric conditions; any point in PC space inside the triangle can be understood as a barycentric combination of the endmembers. By “barycentric” we mean the three non-negative weights, summing to unity, that give the fractional contribution of each triangle vertex to a point inside it. Geometrically, they are the coordinates of the point expressed relative to the triangle’s vertices. For brown dwarfs, these endmembers could represent distinct atmospheric states such as significant cloudiness, temperature extremes, or a hot spot rotating in and out of view. Although these endmembers are not directly observed spectra, they can be reconstructed from the PCs in the same way as the input spectra are decomposed, and therefore represent physically realisable spectral states within the variability sampled by the data. They correspond to limiting cases of the observed atmospheric conditions that could, in principle, be observed if those surface types dominated the visible hemisphere.
Importantly, the endmembers represent only conservative estimates of the distinct surface types and they do not reflect the true surface spectra of any single atmospheric region. Because we observe only disk-integrated spectra (not spatially resolved emission), the observed points in the PCP are themselves mixtures of surface regions. As shown in \citet{Cowan_2013}, the true extreme surface spectra may lie far outside the observed data locus (see their Fig.~2). The endmembers therefore reflect the minimum set of distinct atmospheric states required to explain the variability in our data, but not necessarily the full physical range present on the object.

We further note that our inferred endmembers are specific to this observational epoch. They are defined by the extrema of the observed variability over a single rotation, and therefore depend on both the object and the temporal baseline of the observations. If additional rotations were observed, and if the atmospheric structure evolves on longer timescales, the convex hull, and thus the inferred endmembers, are expected to change as new, more extreme spectral states develop. The endmembers should therefore be interpreted as the most extreme surface types realised during this observation, rather than universal or time-invariant components of the atmosphere.

\begin{figure*}[h!]
    \sidecaption
    \includegraphics[width=12cm]{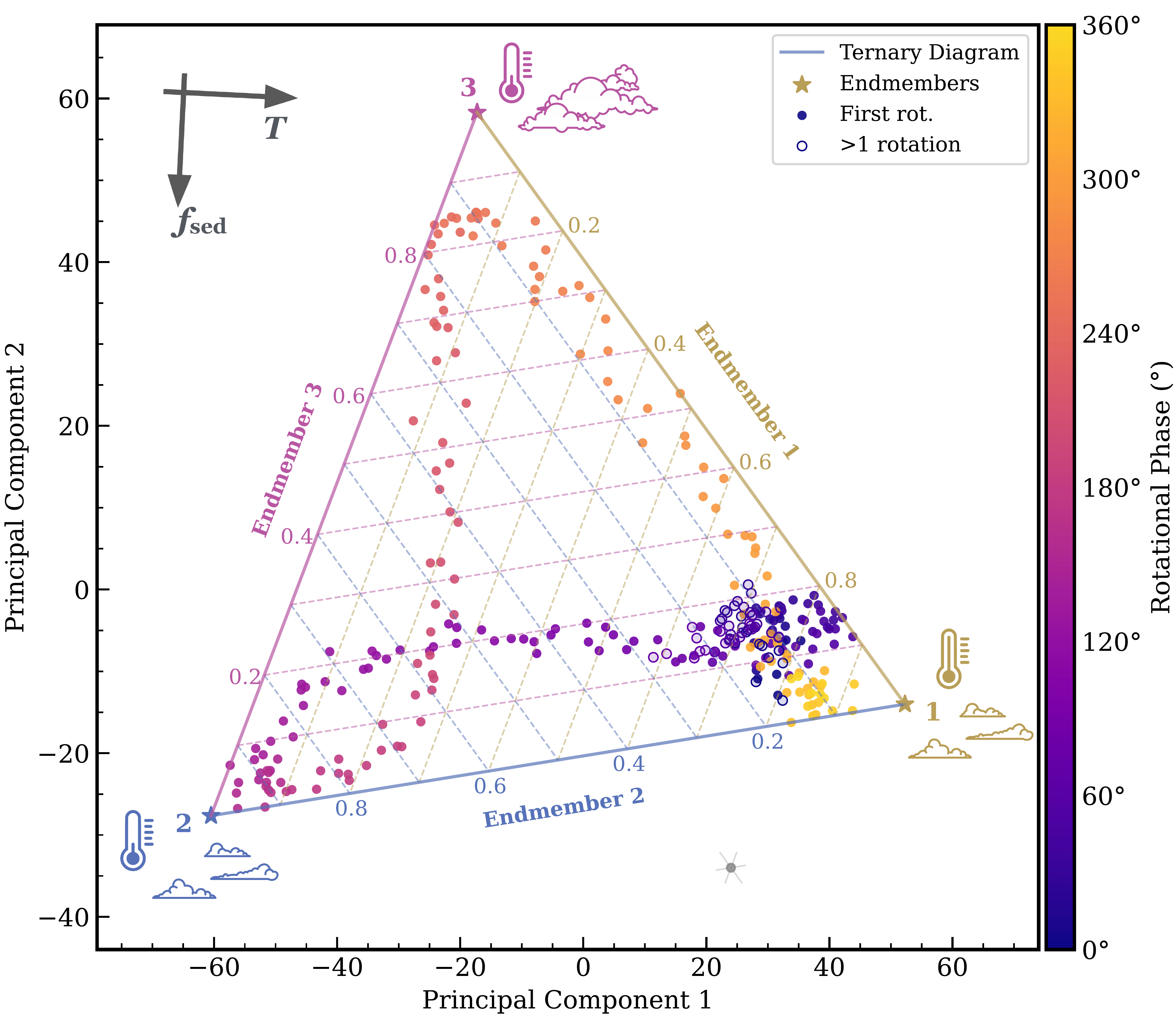}
    \caption{Projection of all time-resolved spectra into the principal-component plane (PCP), coloured by rotational phase.
    The small grey error point near the lower-right illustrates a representative barycentric $1\sigma$ uncertainty.
    The two dark grey arrows annotate qualitative directions inferred from forward-model projections: increasing temperature and increasing cloud sedimentation efficiency $f_{\rm sed}$ (thinner clouds). 
    The inset ternary diagram shows the barycentric coordinates (instantaneous fractional contributions of the three endmembers) as a function of time.
    Stars mark the conservative estimates of the three extreme surface types. The first endmember is characterised by higher temperature and comparatively thinner clouds. The second also exhibits thin clouds but at lower temperature, while the third shows more vertically extended cloud structure at similarly lower temperature.}\label{fig:PCA_tern_surf}
    
\end{figure*}

\subsection{Simplex shrink wrap}

We treated each projection in the plane of the first two PCs as a linear mixture of a small set of underlying “endmember” spectra. We computed the convex hull of all points in the PCP, which describes the smallest convex set that contains the data locus. We then identified the minimum-area triangle that fully encloses the data, following the shrink-wrap procedure of \citet{Cowan_2013} which is congruent with the ternary triangle in Fig.~\ref{fig:PCA_tern_surf}. This construction provides a conservative boundary: only the most extreme observations set the vertex locations. The resulting three vertices (endmembers, marked by stars in Fig.~\ref{fig:PCA_tern_surf}) are spectra that cannot be expressed as linear combinations of others in the dataset and thus serve as estimates of the most distinct atmospheric components sampled over the rotation.

As conservative estimates of extreme spectral surface types, their differences lie below the resolution of the parameter grid in currently available self-consistent forward models. To underline this, the triangle is plotted in red on the model projections in Figure~\ref{fig:MODELS_inPCP}. Comparing the endmembers to the Sonora Diamondback trends in the PCP allows a qualitative interpretation of the modulation. The rotation of \obj{simp} traces a seemingly cyclical trajectory beginning near Endmember~1, associated with higher temperature and relatively large $f_{\mathrm{sed}}$. The data then evolve toward Endmember~2, which shows a modest decrease in temperature while maintaining a similar sedimentation efficiency. Finally, the trajectory approaches Endmember~3, where the temperature is comparable to Endmember~2 but the sedimentation efficiency is significantly lower, implying thicker or more vertically extended clouds. These shifts represent fractional changes in atmospheric parameters.
The relative positioning of the endmembers in model space supports an interpretation of rotational modulation as arising from longitudinal variations in temperature and cloud thickness. 

To visualise the variability represented by the three endmember spectra, we plot their percentage deviation from the time-averaged spectrum in the top panel of Fig.~\ref{fig:PCA_extremes_opacities}. To assess how well these theoretical components are represented by the data, we identified the observed spectra in our time series that lie closest (in Euclidean distance) to each endmember in the PCP. The similarity between these nearest observed spectra and the endmembers, in both spectral shape and deviation from the mean, supports the interpretation that the endmembers represent physically plausible atmospheric states. This comparison is illustrated in Fig.~\ref{fig:PCA_endmember_comp}.

\subsection{Molecular trends and forward-model interpretation}
\label{sec:chem_trends_models}

\begin{figure*}
\centering
\begin{minipage}[t]{.49\textwidth}
    \centering
    \includegraphics[width=0.95\linewidth]{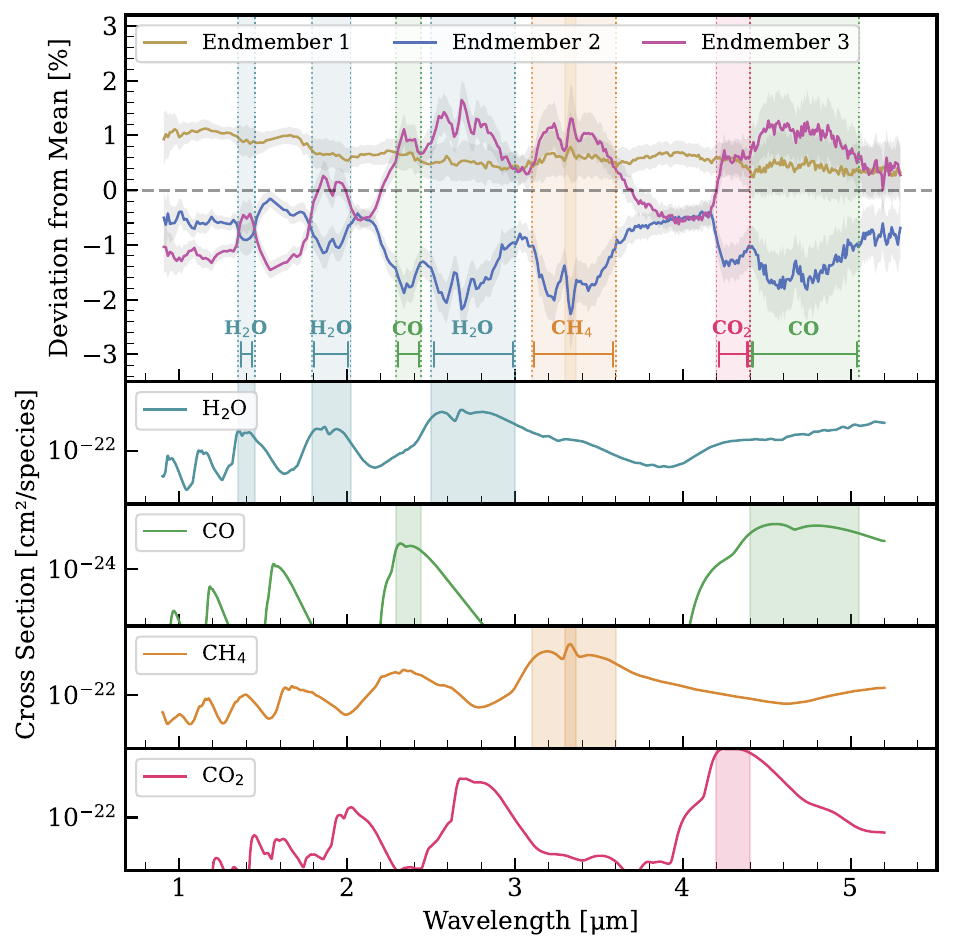}
    \caption{Endmember deviation spectra (coloured; fractional change from the time-averaged spectrum) with gray $1\sigma$ envelopes derived from the nearest observed spectra in PC space (cf. Fig.~\ref{fig:PCA_tern_surf}). Overplotted are PICASO absorption cross sections for H$_2$O, CO, CO$_2$, and CH$_4$. The alignment of molecular bandheads with endmember structure is shaded and highlights chemically sensitive regions.}
    \label{fig:PCA_extremes_opacities}
\end{minipage}\hfill
\begin{minipage}[t]{.49\textwidth}
\centering
    \includegraphics[width=0.95\linewidth]{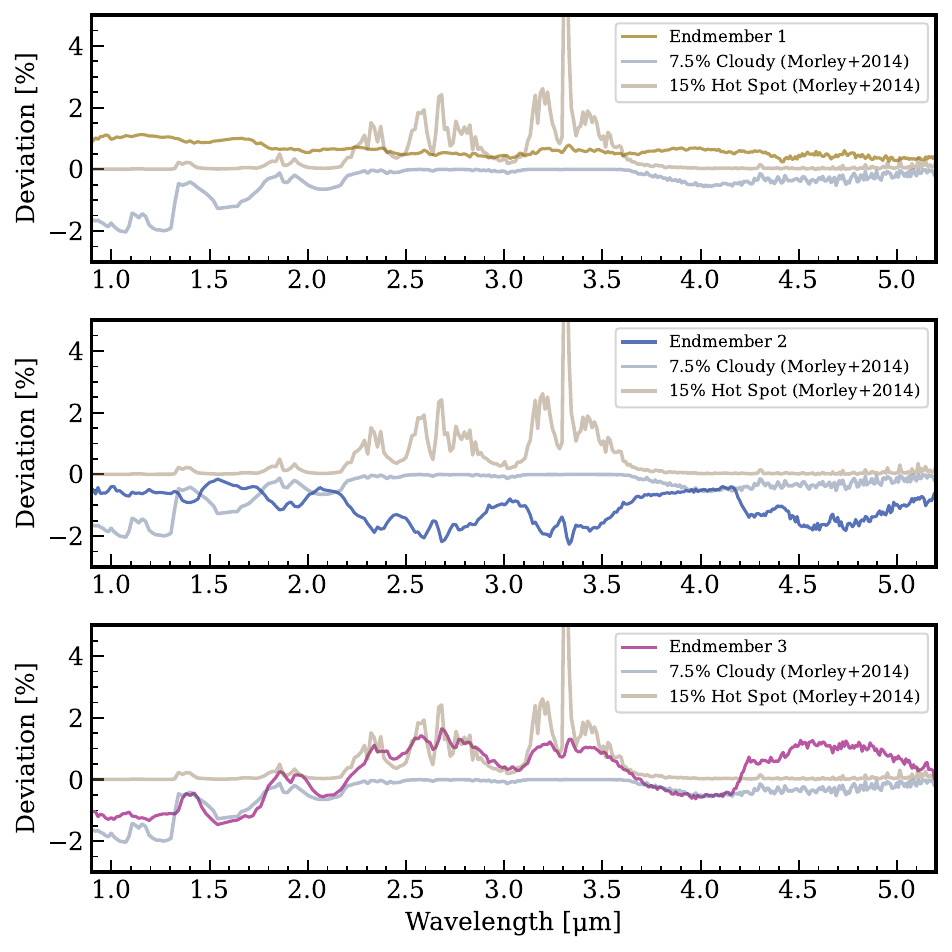}
    \caption{Comparison between the observed endmember deviation spectra (coloured lines) and model predictions from \citet{morley_water_2014} (grey lines). The models simulate fractional spectral deviations resulting from patchy clouds and localised heating at $T\sim1000$\,K.}
    \label{fig:PCA_extremes_morley2014}
\end{minipage}
\end{figure*}

We interpret the endmember deviation spectra by comparing them to molecular cross sections for \htwoo, CO, \chfour, and \cotwo at $T=1100$\,K and $P=10$~bar computed with \texttt{PICASO} \citep{batalha_picaso_2019}, as illustrated by Fig.~\ref{fig:PCA_extremes_opacities}. 
Endmember~1 shows an almost grey offset with only weak chromatic structure, i.e. a broadband brightening relative to the time-mean spectrum without strong reshaping in individual bands. 
Endmember~2 and Endmember~3, by contrast, show opposite behaviour in the major bands: Endmember~2 exhibits deeper \htwoo, \chfour, and CO absorption (more negative deviation from the mean), whereas Endmember~3 shows shallower-than-mean absorption in the same bands (positive deviation in the “deviation-from-mean” plot). 

We also compare the endmembers to perturbations due to (i) varying cloud coverage (blue-gray line in Fig.~\ref{fig:PCA_extremes_morley2014}) and (ii) a localised upper-atmosphere hot spot following \citet{morley_water_2014} (beige line in Fig.~\ref{fig:PCA_extremes_morley2014}). This allows the identification of wavelengths sensitive to a change in cloud cover or upper atmosphere heating. To simulate the variability from these perturbations, we computed fractional deviations from the mean spectrum for mixtures of three underlying components: (a) cloud-free, (b) fully cloudy (silicate + iron condensates), and (c) a hot-spot model with a temperature perturbation at $\sim$0.1~bar. We adopted representative covering fractions (15\,\% cloudy, 7.5\,\% hot spot) and linearly combine the component spectra as in \citet[][their \S2.4]{morley_water_2014}, converting the result to percentage deviation from the mean. We note however, that the models used here are not low gravity. From the models, the $1-2~\mu$m flux windows and the $3.7-4.2~\mu$m region are sensitive to changes in cloud cover, while the hotspot manifests in variability in the  $2.2-3.6~\mu$m region. Both Endmember~2 and Endmember~3 show structure in these cloud- and hot-spot-sensitive ranges. 
In particular, the apparent `negative absorption' of Endmember~2 in the hot-spot-sensitive wavelengths simply indicates that, relative to the time-mean spectrum, the hot-spot contribution is weaker, which reduces the contrast of those features.

The opposite molecular behaviour of Endmembers~2 and~3 may be explained by cloud vertical structure. 
Endmember~3 displays diminished band depths (positive deviations) alongside dimmed cloud-sensitive windows, consistent with thicker, more vertically extended clouds that lift the photosphere to lower pressures and partially mask deeper molecular absorption, thereby muting variability in the bands. 
Endmember~2 shows the reverse trend: stronger band depths (negative deviations) together with brighter windows, consistent with thinner and more settled clouds that permit a deeper view and enhance the contrast between bands and windows. 
Across \cotwo, \htwoo, \chfour, and CO, the endmembers vary coherently, suggesting that changes in cloud vertical extent, rather than large composition changes, set the primary modulation captured by PC2.

The cross-section overlays and the perturbation experiments together indicate that motion in the PCP toward thicker clouds mutes molecular deviations and dims the cloud-sensitive continua, while motion toward thinner clouds brightens the continua and deepens molecular absorption bands. 
This framework provides further interpretation of the endmembers: Endmember~1 represents a broadband luminosity shift with little chromatic change, Endmember~2 is the clearer limit with thinner-clouds and hence enhanced band depths but a dimmer hot spot, and Endmember~3 is the cloudier limit with reduced band depths due to thicker clouds and a brighter hot spot-like emission relative to the mean.

\section{Surface mapping endmember spectra}\label{sec:mapEndmembers}

Following \citet{Cowan_2013}, we interpreted the identified spectral endmembers as conservative estimates of distinct atmospheric states that most strongly modulate the disk-integrated spectrum. Because each observed spectrum mixes light from many longitudes (and latitudes), these endmembers should not be read as “pure” single-region spectra; rather, they are the most extreme combinations present in the data, already blended to some degree by rotational averaging and projection effects. As discussed in \citet[][their Section~5.2]{Cowan_2013}, such conservative surface types limit the range of spectral variability and thus provide a practical basis for mapping.
Our goal is to infer where on the object the dominant spectral states preferentially contribute to the observed spectra. A “surface map” here is a longitudinal brightness pattern for each endmember: a function that specifies how strongly that endmember contributes as a function of longitude, when averaged over latitude and weighted by the visibility of each point on the rotating disk. With these maps, we can relate the time-variable spectral mixing to a simple geometric picture.

Rotational unmixing provides a formal framework for this problem. It describes the inverse problem of recovering both (i) a set of unmixed component spectra and (ii) their time-dependent projected areas from only the rotating, disk-integrated observations \citep{cowan_fuentes_oddHarm_2013}. \citet{Cowan_unphysicalPC_2011} demonstrated that rotational unmixing can recover nearly pure surface spectra from time-resolved, multi-band EPOXI observations of Earth, obtained with the extended \textit{Deep Impact} mission's 30cm diameter telescope coupled with the High Resolution Imager (HRI, \citet{hampton_deepimpact_2005}), and treating Earth as an unresolved exoplanet analogue. However, that success relied on conditions favorable to identifiability, including a small number of independent light curves, relatively low spectral resolution, and very high S/N, such that the number of spatial degrees of freedom was limited. 
Their analysis also benefited from the fact that they observed reflected-light and lacked full phase coverage, producing a narrower convolution kernel and therefore providing stronger spatial constraints on the longitudinal distribution of surface features. In our case, the higher time and wavelength resolution (and the intrinsic complexity of brown-dwarf spectra) introduce more “slices” of the surface rotating in and out of view than can be uniquely constrained by the available, correlated light curves. Given the degeneracies described above, we did not attempt full rotational unmixing. Instead, we fixed the component spectra to the conservative endmembers derived in PCA space, and solved only for their longitudinal distributions (i.e., maps) that best reproduce the observed time dependence.

\subsection{Fractional contribution of endmembers over time}

\begin{figure*}[h!]
    \centering
    \includegraphics[width=0.95\linewidth]{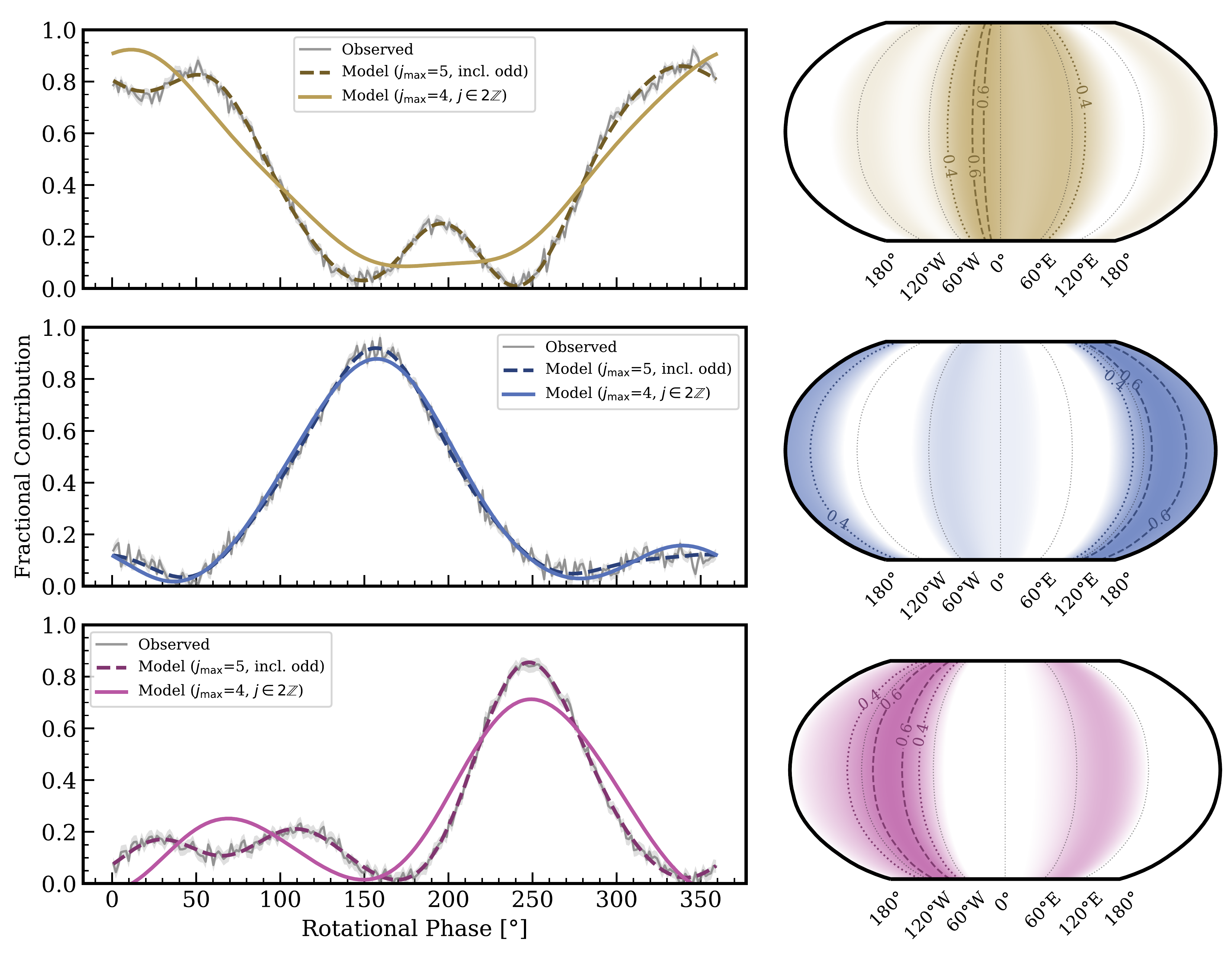}

    \caption{\textit{Left:} Fourier fits to the endmember contribution curves. Solid lines use even-$j$ modes only; dotted lines include all $j$. We adopt $j_{\max}=4$, which minimises the $\chi^2_\nu$; higher orders result in negligible $\chi^2_\nu$ improvement and add unsupported structure. \textit{Right:} Longitudinal surface maps reconstructed from the endmember contribution curves using kernel-corrected Fourier coefficients (even $j$, $j_{\max}=4$). }\label{fig:SURF_fourier+Maps}
\end{figure*}

We approximated each observation as a linear combination of the three endmembers. We visualized this with a ternary diagram (Fig.~\ref{fig:PCA_tern_surf}), in which each point shows the instantaneous fractional weights: points near a vertex are dominated by that endmember; points near the center represent more balanced mixtures.

We estimated uncertainties on these fractional weights by Monte Carlo propagation of the 2D PCA uncertainties into the ternary space. For each spectrum, we took its central location in the PCP and its associated $1\sigma$ uncertainties in PC1 and PC2, and drew $N=100$ random realizations from a Gaussian distribution centred on the measured point. In practice, we assumed the PC1 and PC2 uncertainties are independent and sample from a diagonal 2D normal distribution with standard deviations given by the propagated PC errors. Each realization was then mapped into barycentric coordinates with respect to the shrink-wrapped triangle, and we took the sample mean and standard deviation of the resulting barycentric coordinates as the central values and $1\sigma$ uncertainties on the endmember weights. At the S/N of our data, we found that $N=100$ samples were sufficient for the barycentric means and standard deviations to converge stably.

Reading off the barycentric weights as a function of time results in time series that trace how much of the disk-integrated flux is accounted for by each endmember as the object rotates (contribution curves; grey lines in Fig. ~\ref{fig:SURF_fourier+Maps}). Because our endmembers are conservative (they sit at or near the hull of all observed spectra in PC space), the recovered contributions are lower bounds on the true physical features of the underlying atmospheric states. If the true “pure” extreme spectral states (that we cannot observe due to visibility limitations) lie slightly outside the convex hull spanned by the data, the fitted weights necessarily under-estimate their peak contributions. We therefore assume that the relative phasing and ordering of these contributions is preserved even if the true extreme spectral states lie beyond the hull. This assumption is required to allow further analysis of the data, but it is inherently limited by the fact that the true extreme surface types likely lie outside the shrink-wrapped hull. As shown by \citet{Fujii_trueEndmembers_2017}, allowing endmembers outside the convex hull can rotate the trajectory in the PCP, implying that the directions associated with the true atmospheric states need not coincide with those inferred from the shrink-wrapped endmembers. Consequently, while our approach provides the most conservative interpretation permitted by the observations, the physical representation of the underlying surface types may shift if the true extremes were accessible.

\subsection{Fourier decomposition of endmember contributions}

To turn contribution curves into longitudinal maps, we treated each curve as the disk-integrated flux from a 1D longitudinal brightness pattern and forward-modelled how rotation maps that pattern into phase-dependent flux.

Following the linear forward formalism for rotational light curves, the observed flux is a convolution of the intrinsic map with a non-negative “visibility” kernel that weights only the hemisphere facing the observer \citep[e.g.][]{Cowan_Agol_visibility_2008, cowan_fuentes_oddHarm_2013}. For thermal emission the kernel is simply the cosine of the angle from the sub-observer point, truncated at zero; for an equator-on view this reduces to
$K(\phi-\theta)=\max\!\big[0, \cos(\phi-\theta)\big]$,
where $\phi$ is longitude (increasing eastward) and $\theta$ is rotational phase (sub-observer longitude). This kernel encapsulates foreshortening and the fact that only half the object is visible at any instant, and it is the standard starting point for analytic treatments of rotational mapping. We adopted the equator-on kernel because it imposes a well-defined observability pattern on the harmonics. Inclination primarily mixes spherical-harmonic content; pole-on views suppress rotational modulation (maximal null space), equator-on views maximize recoverable longitudinal structure with a simple kernel. Intermediate inclinations allow limited leakage of odd orders but do not add well-constrained structure; we therefore used the equator-on kernel and reported maps in that limit \citep{Cowan_Agol_visibility_2008, cowan_fujii_summary_2018, Luger_2021}.

We modelled how a 1D brightness pattern $J(\phi)$ (longitude $\phi$) produces a phase-dependent flux. A fully general 2D map could be expanded in spherical harmonics, but with disk integration the visibility kernel strongly suppresses many harmonics and leaves others unconstrained, making inversions ill-posed without making unreasonable assumptions about the surface of the object \citep{cowan_fujii_summary_2018}. We therefore adopted the well-constrained 1D approach of \citet{Cowan_Agol_visibility_2008}. We expanded the longitudinal map in a truncated Fourier series, forward-projected it through a visibility kernel, and fit only those harmonics that project through the kernel (odd harmonics only).

Concretely, we write the longitudinal brightness (for a given endmember) as
\begin{ceqn}
\begin{equation}
J(\phi) \;=\; A_0 \;+\; \sum_{j=1}^{j_{\max}} \Big[ A_j \cos(j\phi) \,+\, B_j \sin(j\phi) \Big],
\end{equation}
\end{ceqn}
where $A_0$ is the mean level and $(A_j,B_j)$ set the amplitudes and phases of longitudinal structure at wavenumber, $j$, (i.e., $j$ bright or dark sectors around the equator). The flux observed at rotational phase $\theta$ is the disk integral of $J$ weighted by the visibility kernel:
\begin{ceqn}
\begin{equation}
F(\theta) \;=\; \int_{0}^{2\pi} J(\phi)\, \max[0,\cos(\phi-\theta)] \, d\phi.
\end{equation}
\end{ceqn}
Because the visibility kernel, $K$, acts as a low-pass filter, only certain combinations of $(A_j,B_j)$ are recoverable from $F(\theta)$. Following \citet{Cowan_Agol_visibility_2008}, we first computed the discrete Fourier coefficients of the measured contribution curve $f(\theta)$,
\begin{ceqn}
\begin{equation}
C_j \;=\; \frac{2}{N}\sum_{k=1}^{N} f(\theta_k)\cos(j\theta_k), \qquad 
D_j \;=\; \frac{2}{N}\sum_{k=1}^{N} f(\theta_k)\sin(j\theta_k),
\end{equation}
\end{ceqn}
and then corrected for the kernel’s suppression using the analytic projection factors, $S_j$. These $S_j$ are the analytic projection factors of the equator-on visibility kernel, i.e., the known transfer function that maps a true longitudinal harmonic into its disk-integrated amplitude; dividing by $S_j$ inverts that filtering for the subset of harmonics that project.
\begin{ceqn}
\begin{equation}
A_j = C_j\,S_j , \qquad B_j = D_j\,(-S_j),
\end{equation}
\end{ceqn}
with
\begin{ceqn}
\begin{equation}
S_j \;=\;
\begin{cases}
\frac{2}{\pi}, & j = 1,\\[4pt]
-\dfrac{(j^2-1)}{2}\,(-1)^{-j/2}, & j>1,\ \text{$j$ even},\\[6pt]
0, & j>1,\ \text{$j$ odd},
\end{cases}
\end{equation}
\end{ceqn}
so that odd $j>1$ terms are null (they do not project into the disk-integrated light curve) and even high-$j$ terms are attenuated. We then reconstructed $J(\phi)$ from the corrected $(A_j,B_j)$ and forward-model $F(\theta)$ for comparison to the observed contribution curves (left panel of Fig.~\ref{fig:SURF_fourier+Maps}). This results in a longitudinal “map” per endmember that shows where on the disk that atmospheric state contributes most strongly. See \citet{cowan_fuentes_oddHarm_2013} for a compact derivation of the thermal visibility kernel, its nullspace structure (e.g. odd $m\!>\!1$ at equator-on), and the inclination dependence of harmonic leakage.

\subsection{Null space and high-j terms} \label{sec:SURF_null_space}

The left panel of Fig.~\ref{fig:SURF_fourier+Maps} shows a comparison of fitting both odd and even modes to the contribution curves (dashed line) or fitting only the even modes and $j=1$ (solid lines).
In the strictly equator-on limit, odd harmonics with with $j>1$ are suppressed entirely by the equator-on visibility kernel and therefore lie in its null space \citep{Cowan_Agol_visibility_2008}: changes in those coefficients do not affect the disk-integrated light curve and cannot be constrained. \obj{simp} is observed at an inclination of $i\sim80^\circ$, rather than exactly equator-on, meaning that structures in the hemisphere tilted away from the observer are visible for only part of a rotation. This partially breaks the north--south symmetry assumed by the equator-on visibility kernel and naturally leaks power into harmonics that would otherwise lie in the null space.

In our fits, including odd $j$'s produced a significantly better match to some features, especially evident in endmembers~1 and 3, indicating that these atmospheric states are likely more strongly north--south asymmetric. The contribution curves themselves also contain small-scale wiggles that are sharper than can be robustly reconstructed with the adopted $j_{\max}=4$ truncation. Some of this structure may reflect inclination-induced leakage into harmonics that would otherwise be suppressed in the strictly equator-on limit. However, because the visibility kernel strongly damps small-scale (high-$j$) structure, the inversion becomes increasingly weakly constrained at high harmonic order. Some of these sharp features may also arise from residual scatter in the principal component trajectory and low-level correlated residuals not fully captured by the formal propagated uncertainties. 
Because the odd terms do not project onto the surface map, we continued to fit only even modes (solid lines in Fig.~\ref{fig:SURF_fourier+Maps}) and show the effect of fitting odd modes on the contribution-curve fits as dashed curves in Fig.~\ref{fig:SURF_fourier+Maps}. 
Comparing models across $j_{\max}$ using reduced $\chi^{2}$ and Bayesian information criteria, we found that $j_{\max}=4$ optimised goodness-of-fit while ensuring we don't over fit (see Fig.~\ref{fig:appendix_jmax_criteria}). Higher orders provided only marginal improvements while introducing unsupported detail (see especially $j_{max}=6$ plot for endmember 1 in Fig.~\ref{fig:appendix_jmax_fit}).

We also note that the adopted kernel neglects wavelength-dependent centre-to-limb variations, including possible limb darkening. A limb-darkened thermal kernel would be more strongly weighted toward disk centre than the simple cosine kernel, and would therefore modify the transfer function that damps high-spatial-frequency structure. In practice, this would change the detailed harmonic suppression factors and hence the precise amplitudes inferred for higher-$j$ terms. However, because our maps are interpreted only at the level of broad longitudinal structure and because the high-$j$ modes are already treated as weakly constrained, we did not attempt to model limb darkening explicitly here.

\subsection{Interpreting endmember surface maps}

We present Robinson-projected maps of the three endmembers in the right panel of Fig.~\ref{fig:SURF_fourier+Maps}. Each map shows broad longitudinal sectors of enhanced contribution, with most of the power in the first two wavenumbers ($j=1,2$). This is expected from the visibility kernel’s low-pass filtering and is consistent with the marginal improvements in fit quality for $j_{\max}>4$ (Appendix Fig.~\ref{fig:appendix_jmax_fit}). The resulting patterns are near-sinusoidal and deliberately free of sharp features; disk integration does not support high-$j$ structure, and our reconstructions avoid over-interpreting unconstrained small-scale contrasts.

Taken together, the three maps imply a longitudinally heterogeneous photosphere in which distinct spectral states rotate in and out of view with modest phase offsets. Read within the PCA framework, as illustrated in Fig.~\ref{fig:PCA_tern_surf}, the first endmember has a higher temperature, resulting in more flux emission (the endmember that has a higher PC1 contribution). Endmembers 2 and 3 differ largely in vertical cloud structure, with similar temperatures, below that of endmember 1. Their preferred longitudes differ, producing the observed phasing in the contribution curves. The longitude of peak contribution for endmember 1 aligns with the longitudinal pattern of brightness variations reported by \citet[][their Fig.~12]{nasedkin_retrieved0136_2025}.

Because the endmembers are conservative, meaning they are extreme spectra identified within the observed convex hull rather than  “pure” surface spectra, the mapped contrasts should be viewed as lower limits on the true longitudinal amplitudes. In practice, we normalise the barycentric weights so that they sum to unity at each epoch; this preserves the relative contributions and phasing of the endmembers even if the true extremes lie outside the data hull. Under the assumption that the unobserved pure spectra would peak near the same longitudes as their conservative proxies (and that longitudinal mixing projects approximately affinely through the kernel), the inferred map geometry (which longitudes favour which spectral state) is robust, while absolute contrasts are conservative. Thus, the maps provide a geometric context for the spectroscopic variability without claiming fine detail that the kernel cannot support.

\section{Comparison with previous PCA studies on \obj{simp}}
\label{sec:comparison}

\subsection{Comparison with \citet{apai_hst_2013}'s PCA on \hst/WFC3 data}
\citet{apai_hst_2013}'s time-resolved HST/WFC3 study of SIMP~0136 found that the observed spectral changes are well reproduced by a linear combination of two distinct spectra, interpreted as a patchy mixture of thick, cooler clouds and thinner, warmer clouds with only modest colour changes. That result is consistent with our finding of a two-dimensional PCP space: both studies point to a small number of spectrally distinct atmospheric states whose changing visibilities drive the observed variability. Our PCA formalism quantifies this with variance fractions and orthogonal eigenspectra, while the HST analysis demonstrated the same low dimensionality via explicit two-spectrum mixing. 

A key difference from the HST/WFC3 analysis of \citet{apai_hst_2013} is that they found a single PC accounted for $\gtrsim 99\,\%$ of the variability, consistent with a two-spectrum mixture in which one surface type dominates exactly as the other disappears. In our \textit{JWST}/PRISM data, the first two PCs together explain $\sim 79\,\%$ of the variance where the PCs contribute in varying ratios to reconstruct the observed variability. This difference likely arises due to the different wavelength ranges of the two instruments. The PCA identifies orthogonal spectral patterns that capture the dominant covariance across the time series, so the number and structure of the recovered PCs depend on which wavelength-dependent variability signals are included in the data. The broader PRISM spectral range (0.9$-$5\,$\mu$m) samples multiple pressure levels \citep{mccarthy_simp_2025}, and includes strong molecular bands (e.g.\ CH$_4$ at 3.3\,$\mu$m, CO$_2$ at 4.3\,$\mu$m) that are absent from WFC3/G141. This allows PCA to separate variability associated with different atmospheric drivers into multiple components. When we restricted the wavelength range of our PRISM dataset to $1.1-1.7$\,$\mu$m and re-ran an unweighted PCA, akin to \citet{apai_hst_2013}'s analysis, $\sim92\,\%$ of the variance collapsed into PC1, with PC2 only capturing $<1\,\%$, reproducing the WFC3-like behaviour. 

\subsection{Comparison with \citet{RomanNIRISS2025}'s PCA on \textit{JWST} NIRISS data}
Time-resolved \textit{JWST}/NIRISS SOSS observations were carried out from UT 05:08:28 to 09:05:07 on 2023 July 22 for GTO program \#1209 (PI: Artigau), 37\,h prior to the NIRSpec/PRISM sequence; see \citet{RomanNIRISS2025}. NIRISS/SOSS provides 81 medium-resolution ($R\!\sim\!1200$) spectra over 2.6\,h, whereas NIRSpec/PRISM covers 2.9\,h with $\sim$5600 integrations at lower resolution ($R\!\sim\!200$). Thus, NIRISS affords finer spectral detail, while NIRSpec offers denser phase sampling across a wider wavelength range.  We analysed both datasets with PCA to explore how the dominant spectroscopic variability drivers may change across epochs. 

To compare the two datasets, we binned NIRISS SOSS' spectral resolution to that of NIRSpec PRISM and adjusted the NIRSpec wavelength range to that of NIRISS (1$-$2.6$~\mu$m). We also median-normalised the NIRSpec dataset to match the NIRISS normalisation.
This normalisation removes achromatic (grey) variability that is common to all wavelengths (e.g. a global level shift or small absolute flux-calibration drifts), so the variance formerly carried by that common mode no longer appears in PC1 and PC2. As a result, the fraction of variance explained by the first two PCs is lower than in the unnormalised NIRSpec-only case.

\begin{figure}[t]
\centering
    \includegraphics[width=0.95\linewidth]{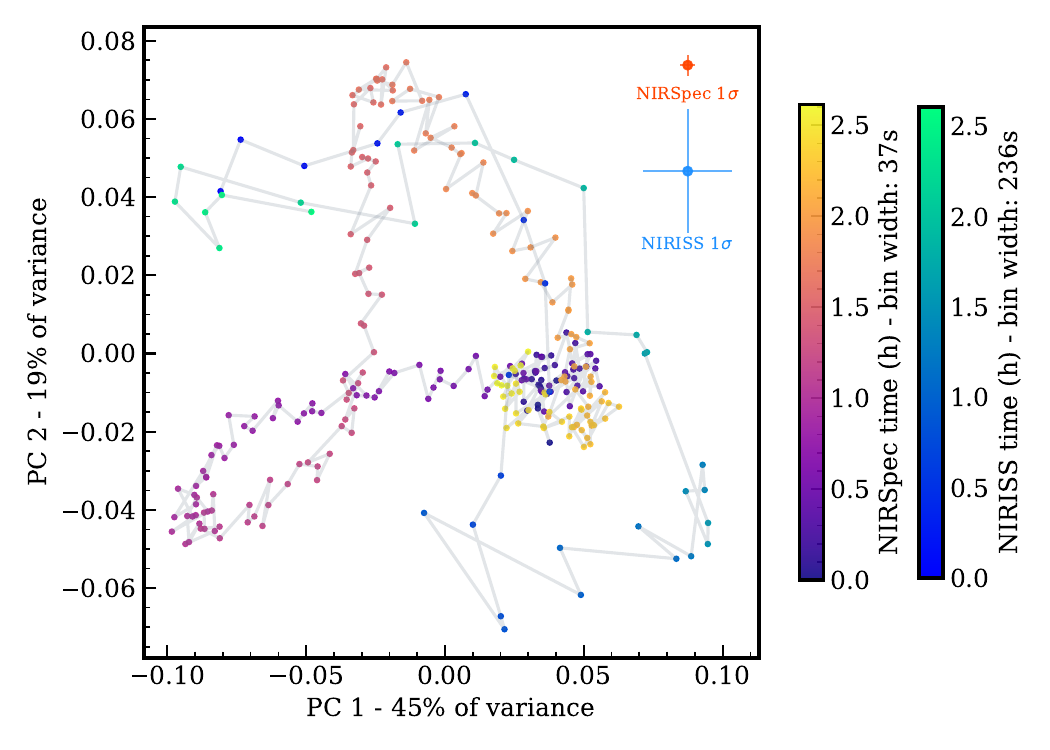}
    \caption{PC1--PC2 projection of PCA trained on the combined dataset of normalised NIRISS (winter colourmap) and NIRSpec (plasma colourmap) spectra with matched wavelength width and binning. NIRSpec forms a coherent loop; NIRISS appears scattered due to low temporal sampling.}
    \label{fig:NIRISSvsPRISM_PCspace}
\end{figure}

For consistency with \citet{RomanNIRISS2025}, and to avoid an instrument-dependent weighting bias, we used an unweighted PCA for the direct NIRISS—NIRSpec comparison in this section. If we applied the same inverse-variance weighting, that we used throughout this manuscript on the NIRSpec data, to the stacked NIRISS+NIRSpec dataset, the higher-S/N NIRSpec spectra would dominate the common PCA basis. In the shared wavelength range, the median per-channel S/N is $\sim115$ for NIRISS and $\sim439$ for NIRSpec, corresponding to median fractional uncertainties of $8.7\times10^{-3}$ and $2.3\times10^{-3}$. Since inverse-variance weights scale as $1/\sigma^2$, NIRSpec would receive a median weight $\sim14$ times larger than NIRISS. The resulting PCA space would therefore primarily reflect the PRISM epoch, with NIRISS projected into a PRISM-dominated basis.

\subsubsection{Comparison in PC space: combined PCA basis}
\label{sec:pcspace_combined}

We first performed PCA on the combined dataset, resulting in one common set of eigenspectra (PC axes) and one common mean spectrum, so that any differences between the two datasets in the PC plane reflect differences in the observed atmospheric state. Figure~\ref{fig:NIRISSvsPRISM_PCspace} shows that both datasets trace a closed, low-dimensional loop, but the loci are not identical in shape or placement within that plane. The NIRSpec sequence draws a smooth, extended loop, whereas the NIRISS points sketch a more compact, sparsely sampled locus.

There are two practical reasons for these differences that do not require different underlying variability mechanisms. Firstly, NIRSpec’s dense sampling resolves the trajectory and captures its full extent while NIRISS, with 81 epochs, samples its trajectory more coarsely. Secondly, because both datasets are mean-normalised to their own respective means, small changes in the amplitude of the relative depth of molecular absorption features compared to nearby continuum regions between epochs will stretch or compress the loop along the same axes. The persistence of a closed, low-dimensional trajectory in the shared PC basis argues that the same small set of quasi-periodic drivers governs both sequences; the geometric imprint (a loop with similar orientation) is stable, while its detailed extent reflects sampling and modest atmospheric evolution. We note that the inherent shapes of the loops differ though, which is likely driven by atmospheric evolution between the two epochs.

Because PRISM time sampling is much finer than that of NIRISS, the combined PCA fit is naturally biased toward the PRISM points. To test for sampling biases, we re-ran the combined analysis after time-binning PRISM to the NIRISS cadence; the shape of the PRISM loop in PC space is relatively unchanged (the eigenvectors shift modestly, as expected when the weighting of time samples changes).  
Empirically, the trajectory (the 2D loop in PC space) is robust to these choices even when the detailed eigenspectra change. We interpret this robustness as evidence that a small number of quasi-periodic drivers with stable phasing govern the variability over the 33.6\,h separating the datasets, while the detailed spectral fingerprints of those drivers evolve with the mean state, wavelength weighting, and normalisation.

\begin{figure}
    \centering
    \includegraphics[width=0.93\linewidth]{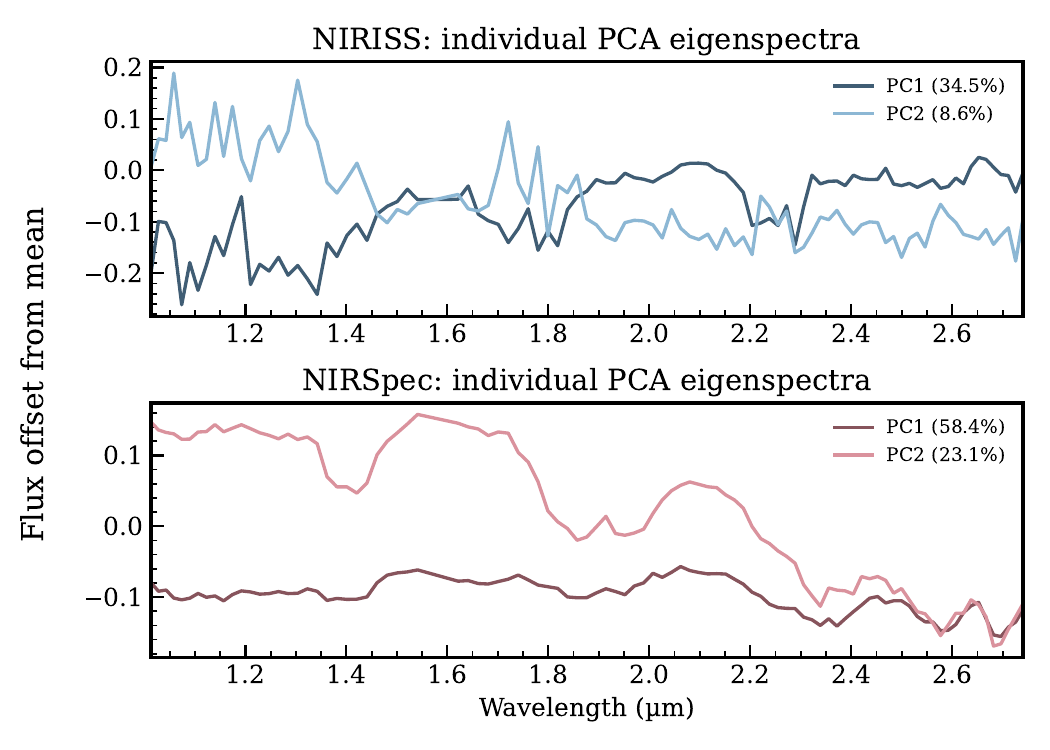}
    \caption{PCs from separate PCA on the NIRISS \textit{(top)} and NIRSpec \textit{(bottom)} datasets, evaluated on the shared wavelength grid and resolution, after median normalisation. PC1 traces temperature in both epochs but their spectral shapes differ.
    Explained-variance fractions are shown in the panel legends.}
    \label{fig:NIRISSvsPRISM_eigenspec}

\end{figure}

\subsubsection{Comparison in PC space: separate PCA fits (per epoch)}
\label{sec:pcspace_separate}

We also ran PCA separately on the NIRISS and NIRSpec datasets (each with its own mean spectrum and covariance), but binned in resolution and clipped in wavelength range as before. In these per-epoch fits the eigenspectra differ at the tens-of-percent level in some wavelength ranges (see Fig.~\ref{fig:NIRISSvsPRISM_eigenspec}), yet when projecting the Sonora Diamondback models into the two PCP planes, the principal axes of variability aligned physically in both cases: the leading component tracks temperature-related changes in flux emission (PC1-like behaviour), and the second component emphasizes molecular bands in a manner consistent with changes in vertical cloud extent ($f_{\rm sed}$). 
This is expected. The PCA decomposition is computed relative to each dataset’s own mean spectrum and wavelength weighting. If the mean state or the relative importance of different wavelengths shifts between epochs (e.g. because cloud opacity or contribution functions change slightly), the eigenspectra will rotate and redistribute power, even when the underlying physics is unchanged. While the exact spectral fingerprints evolve, the dominant physical axes, of temperature for PC1 and band-window contrast tied to cloud structure for PC2, remain stable.

\subsubsection{Same drivers, different fingerprints}

PCA analysis of both datasets points to the same physical picture: two orthogonal drivers capture most of the spectroscopic variability in \obj{simp}, with PC1 tied to temperature-related variability and PC2 emphasising changes consistent with vertical cloud extent ($f_{\rm sed}$) \citep[see also][]{RomanNIRISS2025}. When we ran PCA separately on the two epochs, the principal axes associated with these drivers persist, but the eigenspectra differ. This is consistent with modest evolution of the atmospheric mean state between epochs. This “same drivers, evolving fingerprints” picture echoes multi-epoch results for WISE~J104915.57$-$531906.1AB (Luhman~16AB), where light-curve morphology changes across epochs while the underlying variability mechanisms appear persistent \citep[e.g.][]{chen_1049_2025, biller_weather_2024}. 

As noted previously, the shape of the PRISM loop is stable under different time-binning choices and analysis approaches, indicating that the structure we recovered is not driven by instrumental systematics or reduction choices. This robustness suggests that the difference in loop shape between the NIRISS and NIRSpec observations is physical in origin, pointing to an evolution in the atmospheric state between the two epochs.
Such behaviour is consistent with growing evidence that brown dwarf atmospheres evolve on timescales longer than a single rotation. Long-baseline monitoring has revealed that variability patterns show a continuous evolution in amplitude, phase, and shape over tens to hundreds of rotation periods, driven by changing cloud structures and large-scale atmospheric circulation patterns \citep{fuda_longterm1049_2024}. In particular, complex light curve evolution has been interpreted as arising from planetary-scale waves and evolving cloud systems that modify the observed variability morphology over time. Similarly, multi-epoch observations of VHS~1256~b demonstrate that atmospheric variability can evolve significantly between epochs, further supporting a picture in which brown dwarf atmospheres are intrinsically dynamic and time-variable \citep{Zhou_longterm1256b_2022}.
In this context, the differing loop structures observed here likely reflect genuine temporal evolution in the global cloud distribution and thermal structure.

\section{Summary}\label{sec:summary}

Principal component analysis provides a data-driven framework to describe variability in time-resolved spectroscopy by identifying groups of wavelengths that co-vary over time. These patterns define a set of characteristic spectral shapes, or principal components (PCs), each paired with a time-dependent coefficient that tracks how strongly that pattern contributes at a given rotational phase. PCA expresses the full spectral variability as a combination of a small number of independent spectral components, effectively isolating the dominant physical drivers of variability without requiring prior assumptions about atmospheric structure or surface morphology.

The NIRSpec/PRISM spectroscopic variability of \obj{simp} is well described by the first two weighted PCs, which capture the majority of the variability. After subtracting the first two components, the residual spectra were consistent with the propagated noise floor, with no evidence for additional coherent spectroscopic variability. These PCs therefore provide a compact basis in which the time-resolved spectra can be accurately reconstructed and compared across rotational phase. By projecting Sonora Diamondback models \citep{morley_diamondback_2024} into the two dimensional plane spanned by the PRISM-described PCs (the principal-component plane, PCP), we found that an increase in PC1 correlates primarily with an increase in $T_{\rm eff}$ while motion along PC2 correlates with the cloud sedimentation parameter $f_{\rm sed}$ (cloud vertical extent). Although the Sonora Diamondback models are parameterised by $T_{\rm eff}$, we interpret this not as a change in the intrinsic effective temperature, but as a proxy for atmospheric changes driven by variations in the thermal profile. The PC axes do not align perfectly with the Sonora Diamondback models parametric axes because PCA identifies the orthogonal directions of covariance, whereas the model parameters of $T_{\rm eff}$ and $f_{\rm sed}$ are not mutually uncorrelated in the Sonora Diamondback models. 

This broadly agrees with recent time‐resolved retrievals \citep{nasedkin_retrieved0136_2025}, which find modest ($\sim$5\,K) effective-temperature excursions and no statistically significant phase-coherent evolution in their cloud parameters when all cloud hyperparameters are allowed to vary. In contrast, projections of self-consistent Sonora Diamondback models in the PCP indicate a clear trend along PC2 with the cloud sedimentation parameter $f_{\mathrm{sed}}$, i.e. cloud vertical extent. These two results are not necessarily contradictory since the retrieval framework describes cloud effects with several partially degenerate parameters, which may dilute the phase trend of any one parameter even when the net cloud effect varies, while the grid projection compresses cloud physics largely onto a single parameter ($f_{\mathrm{sed}}$). In summary, the dominant variability is set by changes in temperature (PC1) in combination with vertical-cloud-extent modulation of spectral bands (PC2).

The low-dimensional structure of the PCP also implies that the observed spectra can be described as mixtures of a small number of distinct atmospheric states. In two dimensions, the smallest simplex that encloses the data is a triangle whose three vertices provide conservative estimates of the most extreme spectral states sampled over the rotation. These endmembers should not be interpreted as spatially pure surface spectra. Because the observations are disk-integrated, even the most extreme points in the PCP are themselves mixtures of multiple surface regions. Instead, they define the minimum set of distinct atmospheric states required to reproduce the observed variability. Their relative positions in the PCP indicate that the first endmember corresponds to a hotter and relatively thinner-cloud state, while the other two occupy a cooler area of the PCP that differs primarily in cloud vertical extent.
By following the barycentric coordinates of the observed spectra within this triangle, we traced how the relative contributions of these three atmospheric states vary with rotational phase. This shows that the spectroscopic variability of \obj{simp} is well described as a changing mixture of three physically distinct and phase-separated states rotating in and out of view. The resulting longitudinal maps recover broad sectors of enhanced contribution with modest phase offsets between the endmembers, indicating a longitudinally heterogenous photosphere. The improved fit obtained when odd Fourier modes are allowed in the endmember contribution curves suggests that the atmosphere is not strictly north--south symmetric, consistent with previous inferences from independent \textit{JWST}/NIRISS analyses \citep{RomanNIRISS2025}.

Comparison with these NIRISS/SOSS observations of the same object, obtained 37\,h earlier \citep{RomanNIRISS2025}, shows that the same physical drivers persist across epochs, even though their detailed spectral fingerprints evolve. After matching wavelength coverage and normalisation, both datasets trace a closed loop in principal-component space, indicating that the variability is governed by the same small set of quasi-periodic atmospheric processes. When analysed separately, the eigenspectra differ in detail. This is likely due to the evolution of the mean atmospheric state and the pressure levels that contribute to the emergent flux. Performing the same projection of the Sonora Diamondback models for each dataset independently confirmed that the physical interpretation of the two dominant axes remains unchanged: a temperature-like component and a cloud-structure component. \obj{simp}'s atmosphere evolves within a stable low-dimensional variability manifold, while the detailed spectral trace of that variability evolves from epoch to epoch. Continuous long-term observations are required to robustly characterise the nature of this evolution.

\section{Conclusions}\label{sec:conclusions}

The existence of two dominant variability components implies that the spectra of \obj{simp} can be described as mixtures of three distinct atmospheric states. These states represent spatially separated regions on the object that rotate in and out of view and combine linearly to produce the observed spectra. This provides a physically intuitive picture of the variability wherein \obj{simp}'s atmosphere is heterogeneous and composed of a small number of recurring spectral states whose changing visibility governs the observed spectra, rather than variability arising from fully stochastic or high-dimensional processes. The fact that these states peak at different rotational phases indicates that they occupy different longitudes, supporting a patchy atmospheric structure that cannot be described by a single 1D forward model at any given time.

The directions of variability identified directly from the data align with those traced by self-consistent forward models. The PCs derived from the observations capture a large fraction of the variance across the Sonora Diamondback model grid, indicating that the same physical processes that differentiate models in parameter space also govern the temporal variability of the atmosphere. This shows that forward models do not merely describe static atmospheric states, but capture the way in which atmospheres vary. Even with a coarse grid, the physical processes that drive differences between models appear to be the same physics driving the observed variability.

The PCA-derived interpretation is consistent with previous analyses of this dataset, including time-resolved retrievals and variability studies \citep[e.g.][]{nasedkin_retrieved0136_2025, mccarthy_simp_2025}. The PC analysis recovered the same physical picture, but in a fully data-driven and computationally efficient way. This makes it a natural first step for analysing time-resolved spectroscopy by isolating the dominant drivers of variability and identifying the phases where they are most strongly expressed, providing a way to select epochs for detailed retrieval analyses. 

The PCA results also offer an explanation for the wavelength-dependent phase lags reported in previous multi-band monitoring. Broad-band light curves can be expressed as different linear combinations of a small number of underlying variability modes, rather than requiring all wavelengths to trace the same atmospheric structure with a fixed phase offset. Bandpasses dominated by the PC1-like component will therefore exhibit different light-curve morphologies from those more sensitive to the PC2-like component, with intermediate wavelength regions showing intermediate behaviour. Under this interpretation, the large phase offsets observed between \hst/WFC3 and \spitzer/IRAC light curves \citep[e.g.][]{yang_extrasolarstorms_2016, buenzli_vertical_2012} arise from differing contributions of distinct pressure-sensitive variability modes, rather than from a single atmospheric structure viewed with a wavelength-dependent delay.

This method is immediately applicable to the growing set of \textit{JWST} time-resolved observations across a wide range of atmospheric regimes. Extending this analysis to longer-wavelength \textit{JWST}/MIRI datasets will probe higher atmospheric layers and complementary chemical species, while application to hotter objects such as WISE~J104915.57$-$531906.1A \citep[L7.5,][]{burgasser_1049A_2013} and cooler targets such as Ross~458c \citep[T8,][]{burgasser_clouds_2010} will test how the dominant modes of variability evolve across a range of temperatures, gravities, and cloud regimes. Applying PCA consistently across these datasets will allow direct comparisons of the variability structure between objects and establish whether the low-dimensional behaviour observed in \obj{simp} is common in substellar atmospheres.

The differences in \obj{simp}’s atmosphere between the NIRSpec/PRISM and NIRISS/SOSS observations, which were separated by $33.6$\,h or $13.9\pm0.5$ rotations demonstrate that single-rotation datasets, while providing a complete longitudinal snapshot, capture only one state of the atmosphere. In this sense, a single rotation gives a 360$^\circ$ view of the object but does not capture how the object evolves from one of these states to the next. The reshaping of the PCP loci across the two observations indicates that the underlying atmospheric state evolves on timescales of tens of rotations, but the nature of this evolution remains unconstrained. Continuous, multi-rotation observations would allow this evolution to be tracked directly in principal-component space: whether the atmosphere repeatedly traces a stable, closed trajectory or shifts across the plane to new regions corresponding to different global states. A stable, repeating trajectory would suggest persistent atmospheric structures whose relative contributions vary with rotation, while systematic shifts in position or shape would indicate global changes in flux and cloud properties. The manner of this evolution is equally relevant: gradual transitions between trajectories would be consistent with slowly evolving, radiatively driven structures such as banded flows or long-lived spots, where cloud feedback and thermal forcing organise the atmosphere into quasi-stable patterns \citep[e.g.][]{apai_zonesbands_2017}. Rapid changes or the presence of multiple intermediate trajectories would point to a more turbulent, dynamically evolving atmosphere in which intermittent forcing, strong radiative--convective coupling, or localised heating and cooling drive continual reconfiguration of the emitting layers \citep[e.g.][]{tan_atmospheric_2021, Lee_h2s_2024}. Studying how these atmospheres transition from one weather state to the next is therefore essential for understanding the climate and underlying dynamics of these extrasolar worlds.

\begin{acknowledgements}
    The authors would like to thank the anonymous reviewer for their thorough and  helpful feedback. MAS acknowledges support from Trinity College Dublin via a  Trinity Research Doctorate Award. MAS acknowledges support from Georg Sekundus Schrader and Charlotte-Grace Touchard-Paxton. EN, JMV and CO'T acknowledge support from Royal Society - Research Ireland University Research Fellowship (URF/1/221932, RF/ERE/221108). JMV, AMM and SB acknowledge support from the European Union through the Exo-PEA ERC project (grant number 101164652). Views and opinions expressed are, however, those of the author(s) only and do not necessarily reflect those of the European Union or the European Research Council Executive Agency. Neither the European Union nor the granting authority can be held responsible for them.
\end{acknowledgements}

\bibliographystyle{yahapj} 
\bibliography{bib}

\begin{appendix}
\section{Model projections into the PCP}\label{sec:app_modelproj}

We projected the Sonora Diamondback models into the PCP derived from \obj{simp}'s variability. Since the PCA basis used in this work was defined after noise whitening, the models were processed with the same mean subtraction and wavelength-dependent uncertainty scaling as the observed spectra before projection. We assembled a matrix of model spectra 
$\mathbf{M}\in\mathbb{R}^{N_{\rm mod}\times p}$, where each row corresponds to a single model spectrum after binning and scaling to the observed wavelength grid, and each column corresponds to a wavelength channel.

We used the PCA basis derived from the observations (Eq.~\ref{eqn:SVD_1}), characterised by the mean spectrum $\bar{\mathbf{x}}$, the wavelength-dependent uncertainty scale $\mathbf{s}$, and the matrix of eigenspectra $V \in \mathbb{R}^{p \times p}$. Let $V_K \in \mathbb{R}^{p \times K}$ denote the first $K$ PCs, i.e.\ the first $K$ columns of $V$. Each model spectrum was first mean-subtracted using the observational mean and then divided by the same uncertainty scale used to whiten the data,
\begin{ceqn}
\begin{equation}
\mathbf{M}' =
\left(
\mathbf{M} - \mathbf{1}\bar{\mathbf{x}}^{\!\top}
\right)
\oslash
\mathbf{s}^{\!\top},
\end{equation}
\end{ceqn}
where $\mathbf{1}\in\mathbb{R}^{N_{\rm mod}}$ is a column vector of ones, $\oslash$ denotes element-wise division, and $\mathbf{s}$ is the vector of wavelength-dependent uncertainty scales. The model spectra were then projected into the principal-component space as
\begin{ceqn}
\begin{equation}
\mathbf{Z}_K = \mathbf{M}' V_K.
\end{equation}
\end{ceqn}
The matrix $\mathbf{Z}_K\in\mathbb{R}^{N_{\rm mod}\times K}$ contains the PC coordinates for each model spectrum along the first $K$ PCs. These coordinates are directly analogous to the PC scores of the observed spectra and are expressed in the same noise-whitened PCA space.

We reconstructed the model spectra using only the first $K$ components by first reconstructing in the whitened space and then transforming back into flux units,
\begin{ceqn}
\begin{equation}
\widehat{\mathbf{M}}_K =
\left(
\mathbf{Z}_K V_K^{\!\top}
\right)
\odot
\mathbf{s}^{\!\top}
+
\mathbf{1}\bar{\mathbf{x}}^{\!\top},
\end{equation}
\end{ceqn}
where $\odot$ denotes element-wise multiplication. The matrix $\widehat{\mathbf{M}}_K$ is therefore the approximation of the model spectra in flux space using a truncated PCA basis, consistent with the reconstruction in Eq.~\ref{eqn:weighted_reconstruct}.

The fraction of variance in the model grid reconstructed by the first $K$ PCs is then
\begin{ceqn}
\begin{equation}
R_K =
1 -
\dfrac{
\sum_{j=1}^{p}
\mathrm{Var}
\!\left(
\mathbf{M}_{:,j}
-
\widehat{\mathbf{M}}_{K,:,j}
\right)
}
{
\sum_{j=1}^{p}
\mathrm{Var}
\!\left(
\mathbf{M}_{:,j}
\right)
}\,,
\end{equation}
\end{ceqn}
where $\mathbf{M}_{:,j}$ denotes the flux values of all models at wavelength index $j$, and the variance is taken across the model ensemble. The numerator represents the residual variance after reconstruction in flux space, while the denominator is the total variance in the model grid. This quantity therefore measures how well the observationally derived PCA basis reconstructs the spectral diversity of the forward-model grid.

\section{Comparison of Endmember deviation spectra to closest observed spectra}\label{appendix_endmember_comp_to_obs}

\begin{figure}[t!]
  \centering
  \includegraphics[width=0.45\textwidth]{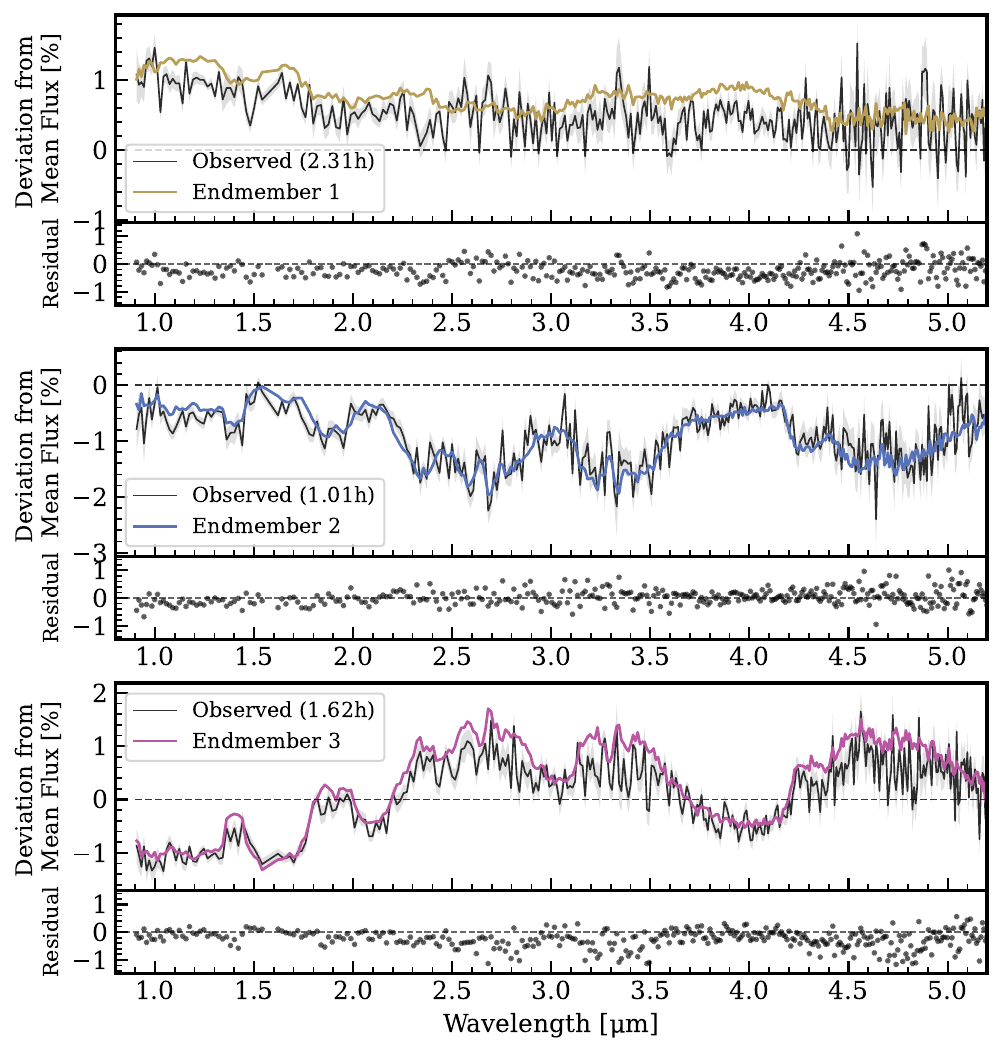}
  \caption{Endmember deviation spectra (solid lines) compared to the nearest observed counterparts (dotted lines), plotted as percentage deviations from the time-mean spectrum. For each endmember, the “nearest observed” spectrum is the epoch whose barycentric weight for that endmember is maximal. Residuals (observed minus endmember) show the observed spectrum's deviation from a perfectly “pure” surface type. Faint shaded bands indicate the $1\sigma$ uncertainty from observational errors and the Monte-Carlo uncertainty in the endmember location.}
  \label{fig:PCA_endmember_comp}
\end{figure}

To assess how the PCA-derived endmembers relate to the observed data, we identified the closest observed spectrum to each vertex of the triangle in the principal-component plane, by minimising the Euclidean distance in the PC1--PC2 plane. This distance was measured in the same PCA coordinates used throughout the analysis. We then compared these spectra to the corresponding endmember reconstructions in Fig.~\ref{fig:PCA_endmember_comp}. The endmember spectra were reconstructed by transforming their PC-space coordinates back into flux space using the inverse PCA reconstruction described in Eq.~\ref{eqn:weighted_reconstruct}. This provides a direct empirical comparison for the endmembers' deviation from the mean spectra, demonstrating that they correspond closely to real, observed atmospheric states. The reconstructed endmember spectra appear smoother than the data, as they are computed using only the first two PCs and therefore capture only the dominant modes of variability. As such, they represent idealised, extreme surface types rather than fully resolved spectra, but nonetheless retain the key spectral characteristics of the closest observed states.

\section{Retrieved model projections in the PCP: CO, \chfour, \htwos}\label{sec:app_retrievals}

\begin{figure*}[ht] 
  \label{ fig7} 
  \begin{minipage}[b]{0.5\linewidth}
    \centering
    \includegraphics[width=0.95\textwidth]{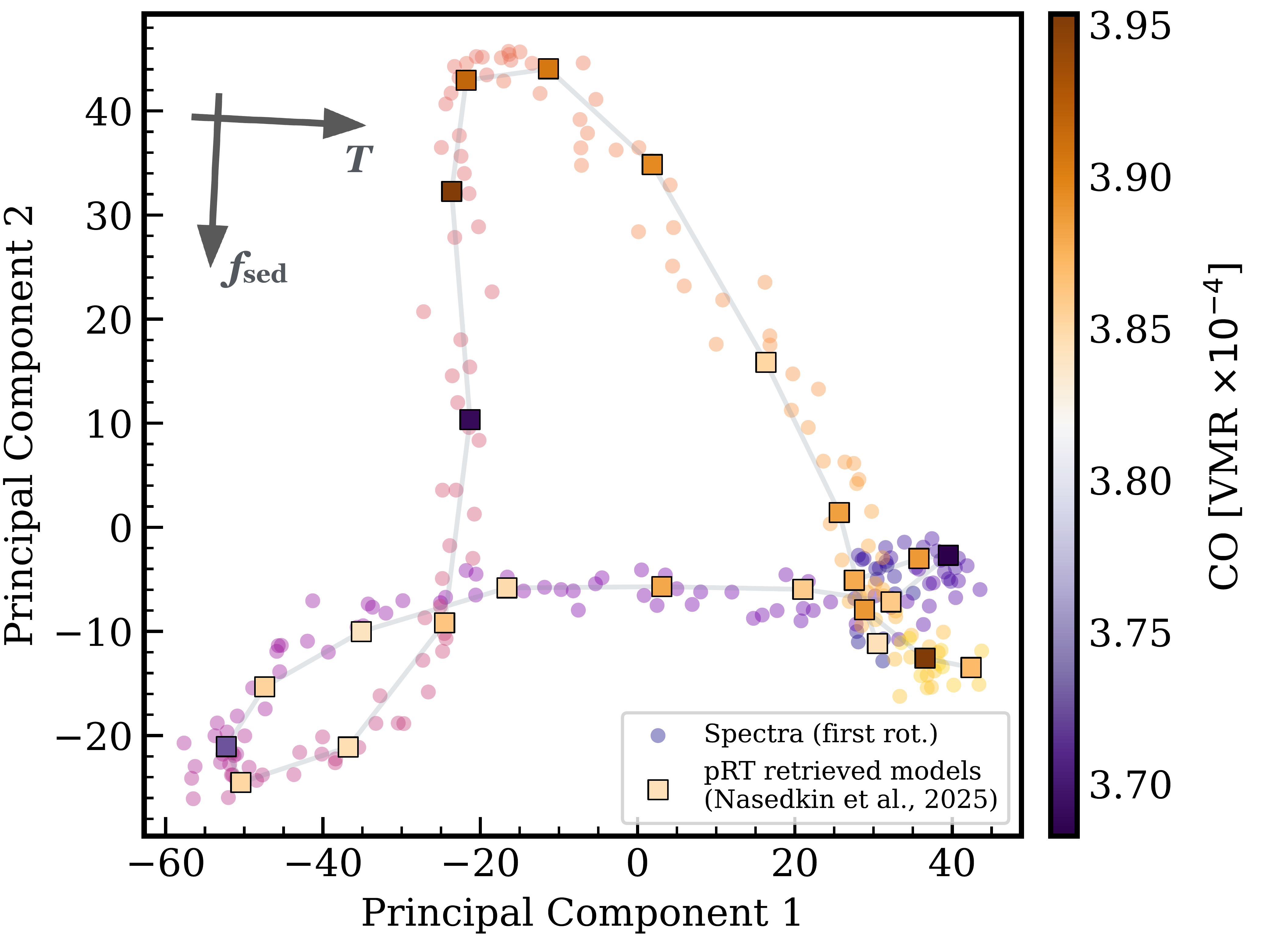}
    \vspace{2ex}
  \end{minipage}
  \begin{minipage}[b]{0.5\linewidth}
    \centering
    \includegraphics[width=0.95\textwidth]{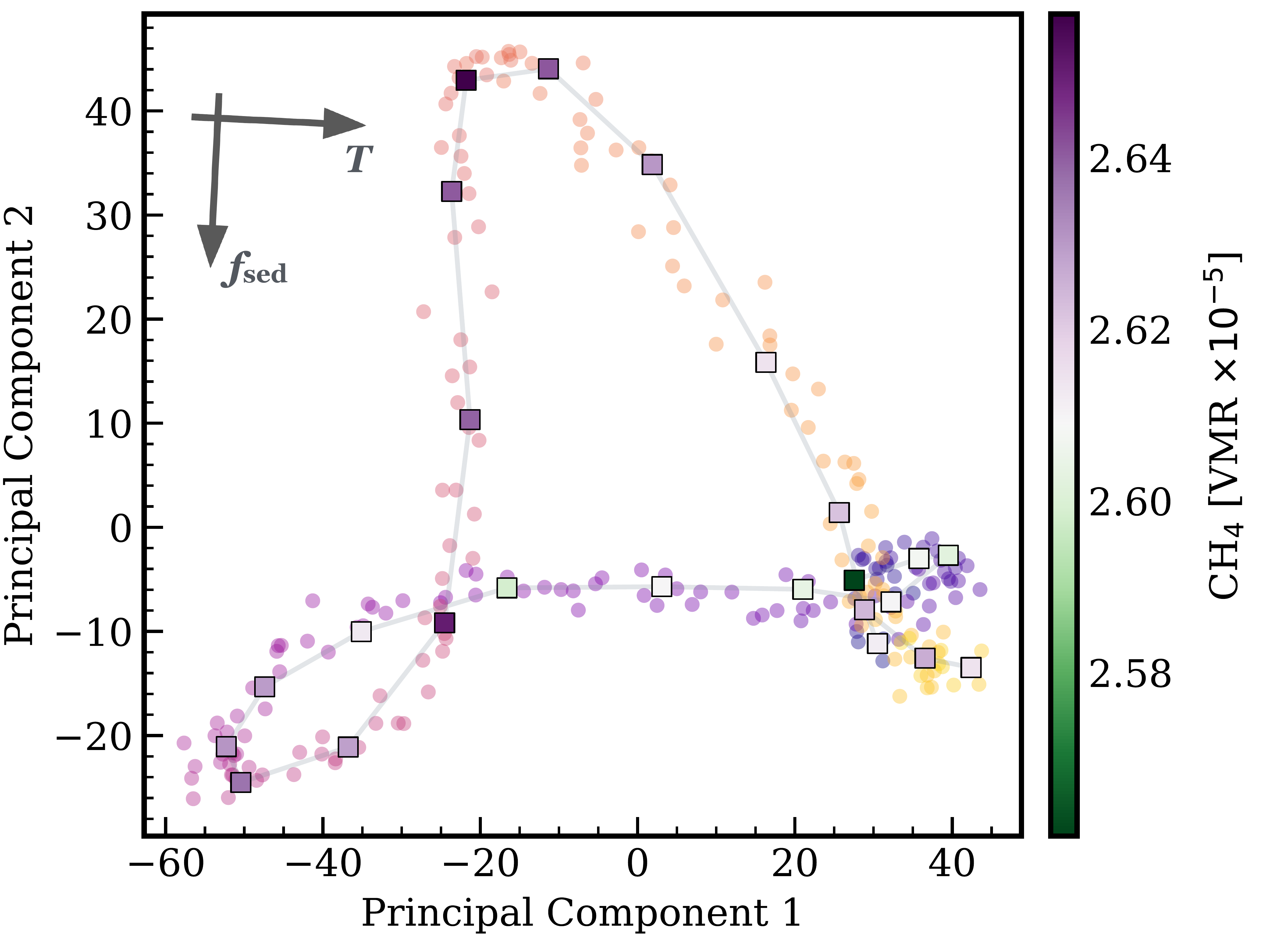}
    \vspace{2ex}
  \end{minipage} 
  \begin{minipage}[b]{0.5\linewidth}
    \centering
    \includegraphics[width=0.95\textwidth]{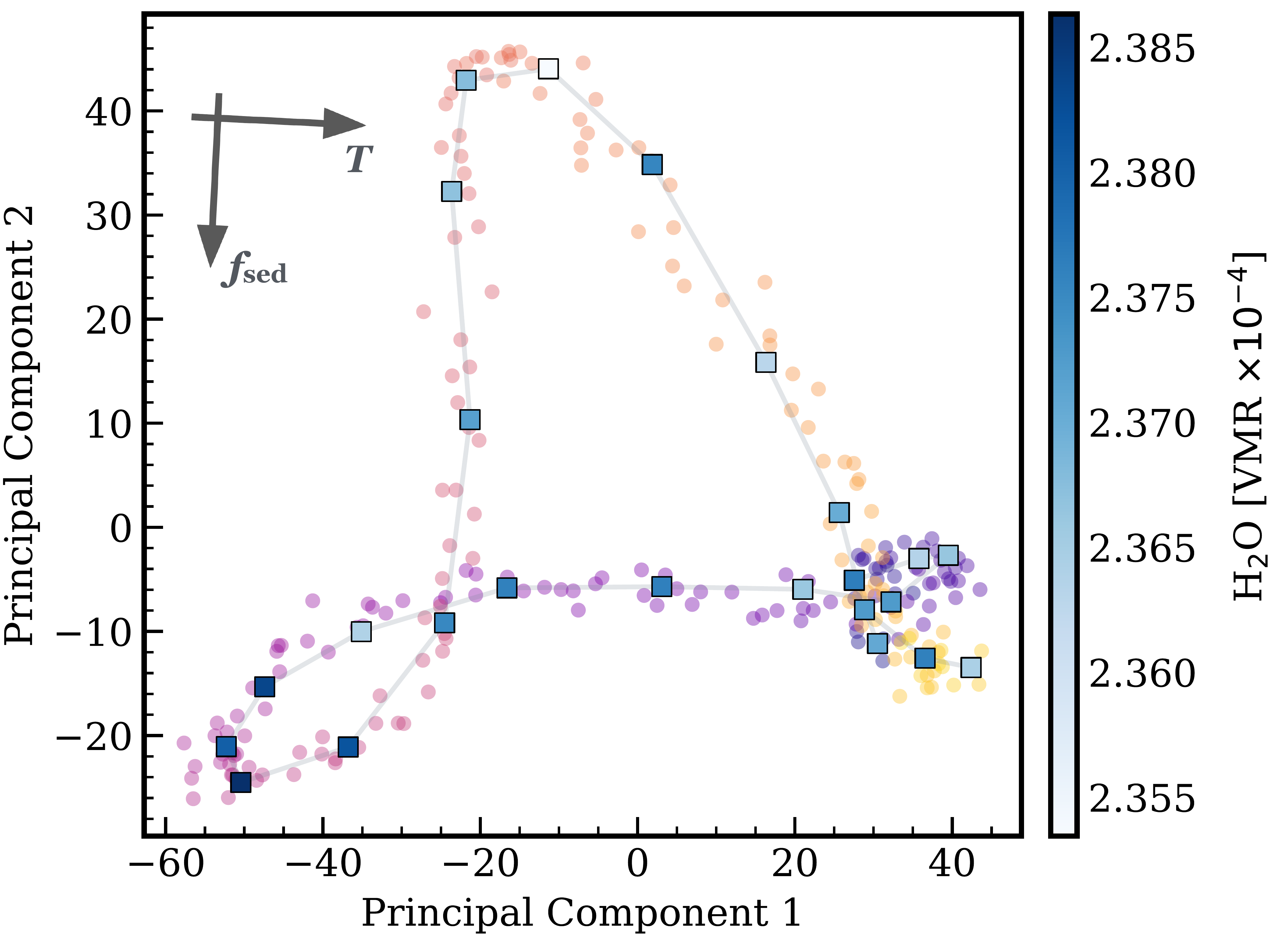}
    \vspace{2ex}
  \end{minipage}
  \begin{minipage}[b]{0.5\linewidth}
    \centering
    \includegraphics[width=0.95\textwidth]{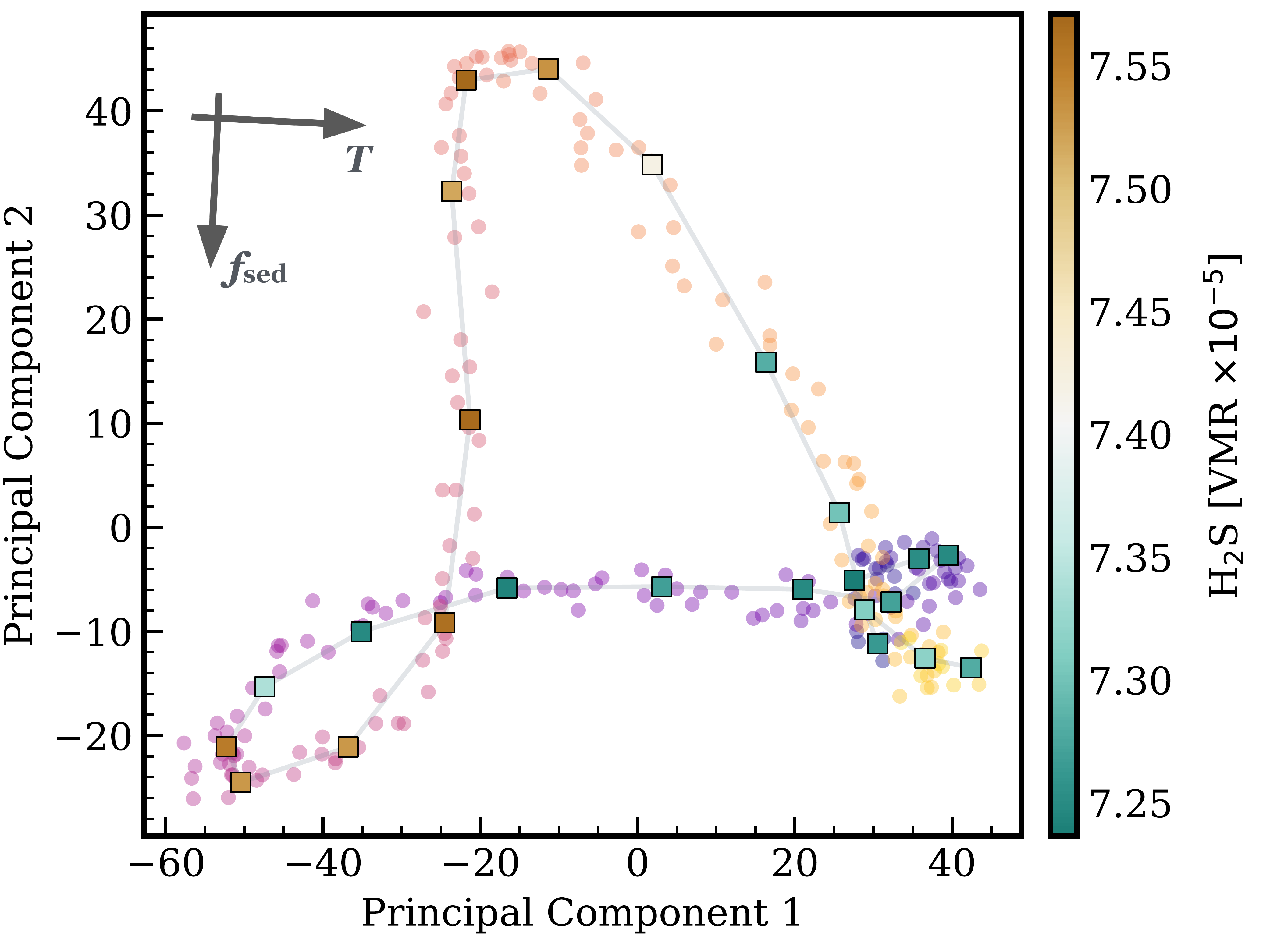}
    \vspace{2ex}
  \end{minipage} 
  \caption{Parameters retrieved by \citet{nasedkin_retrieved0136_2025} projected into the PCP, coloured by: \textit{Top left:} CO abundance, \textit{Top right:} \chfour abundance, \textit{Bottom left:} \htwoo abundance and \textit{Bottom right:} \htwos abundance. We found no clear trend with either PC or model-projected axes for CO or \chfour. We found a tentative decreasing trend in \htwoo abundance with increasing PC2. We found abrupt changes in \htwos abundance near corner points of the data locus.}\label{fig:appendix_retr_PCA}
\end{figure*}
The retrieval framework of \citet{nasedkin_retrieved0136_2025} allows for vertically varying abundance profiles for certain species, including CO and \chfour, rather than enforcing a constant mixing ratio with pressure. As a result, the inferred abundance of these species depends on the pressure level at which it is evaluated. For comparison across rotational phase and for projection into principal-component space, we therefore required a representative abundance corresponding to the photospheric layers probed by the observations.
We defined this representative abundance as the median volume mixing ratio within the photosphere for each retrieved spectrum. The photospheric region was approximated by a characteristic pressure level near $P \sim 0.1$~bar, consistent with the pressures probed by the NIRSpec/PRISM wavelength range. However, the native pressure grid of the retrieval output is relatively coarse, and does not always include a grid point exactly at this pressure. In some cases, the closest grid points (e.g.\ $\sim$0.08 and $\sim$1.6~bar) span a range over which the abundance profile can vary significantly, particularly for species with strong vertical gradients.
To obtain a consistent estimate of the abundance at a fixed pressure level, we interpolated the retrieved abundance profiles in log-pressure space. Specifically, for each phase-resolved retrieval, we interpolated the CO and \chfour\ mixing ratio profiles to a common reference pressure of $P = 0.093$~bar, chosen as the location where the profiles are most consistently sampled across all retrievals. 
The resulting abundances represent the median volume mixing ratios of CO and \chfour\ at the chosen photospheric pressure for each of the 24 phase-resolved retrievals. These values were used in these appendix figures to assess whether coherent phase-dependent trends are present in the carbon chemistry, and to compare their behaviour with that of \cotwo\ in principal-component space.

We did not find a clear, coherent phase-dependent trend in either the CO or \chfour\ abundances. This is consistent with the findings of \citet{nasedkin_retrieved0136_2025}, who report that neither species exhibits statistically significant variability across the rotation. While CO and \chfour\ are expected to be linked through carbon chemistry and disequilibrium processes, the absence of a strong, phase-coherent signal in these species may reflect limitations in sensitivity at PRISM resolution, degeneracies within the retrieval framework, or the fact that their variability is smaller than that of other parameters such as \cotwo.

We also examined the behaviour of \htwoo, which is a dominant opacity source across the near- and mid-infrared and therefore plays a key role in shaping the emergent spectrum. Its retrieved abundance shows a trend of decreasing abundance with increasing PC2 across the PCP. However, \citet{nasedkin_retrieved0136_2025} report no statistically significant phase trend. Given its central role in both radiative transfer and carbon–oxygen chemistry (e.g.\ through reactions linking CO, \chfour, and \cotwo), even modest variations in \htwoo\ could influence the apparent behaviour of other species, although we did not find strong evidence for coherent variability in this case.

In contrast, \htwos\ exhibits statistically significant variability in the retrievals \citep{nasedkin_retrieved0136_2025} and shows a distinct, non-monotonic pattern in principal-component space. Its abundance increases sharply near the joint minimum of PC1 and PC2, followed by elevated values over a range of phases before declining again. This behaviour does not map cleanly onto either of the dominant PCA axes, suggesting sensitivity to more localised or transient atmospheric processes. One possible interpretation is that \htwos\ traces spatially confined disturbances (e.g.\ storm-like features) that modify the local thermal structure and chemical partitioning over limited longitudinal regions.
At the same time, we caution that \htwos\ may act as a compensating species within the retrieval framework. In the NIRSpec/PRISM wavelength range, \htwos\ contributes opacity over relatively broad spectral regions without strong, uniquely identifiable features, making it susceptible to degeneracies with cloud opacity, temperature structure, or other absorbers. As a result, some of the apparent variability in \htwos\ may reflect the retrieval adjusting this parameter to account for residual mismatches in the spectral fit. A more detailed discussion of the possible physical and retrieval-driven interpretations of the \htwos\ variability is provided in this appendix.

\section{AIC and $\chi^2_\nu$ of $j$ fits}\label{sec:app_BIC_jSurfMaps}
To set the maximum harmonic included in the longitudinal mapping (Sec.~\ref{sec:SURF_null_space}) we compared even-only models (always including $j{=}1$) across truncation orders $j_{\max}=\{2,4,6,8,\ldots\}$ using both the reduced $\chi^2$ and the Akaike Information Criterion (AIC) \citep{akaike1974new}. The reduced $\chi^2$ curves exhibit shallow minima at $j_{\max}{=}4$ for Endmembers~1 and~3, while Endmember~2 shows a slightly lower minimum at $j_{\max}{=}6$. Although this could be taken to favour higher spatial detail for Endmember~2, we interpret the gain as marginal and likely unsupported by the visibility kernel (which strongly damps high-$j$ structure). For physical parsimony and consistency across components, we therefore adopt $j_{\max}{=}4$ for the baseline maps.

By contrast, the AIC typically prefers $j_{\max}{=}2$. This is expected: AIC penalizes added parameters ($+2k$), so when the improvement in fit quality at higher $j_{\max}$ is small—because kernel smoothing leaves little high-order power—the penalty outweighs the modest drop in $\chi^2$. Visually, the $j_{\max}{=}2$ and $j_{\max}{=}4$ surface maps are very similar, underscoring that most of the information resides in the first few harmonics. We retained $j_{\max}{=}4$ to allow a conservative amount of structure while avoiding over-interpretation of high-order terms.

\begin{figure}[h]
  \centering
  \includegraphics[width=0.45\textwidth]{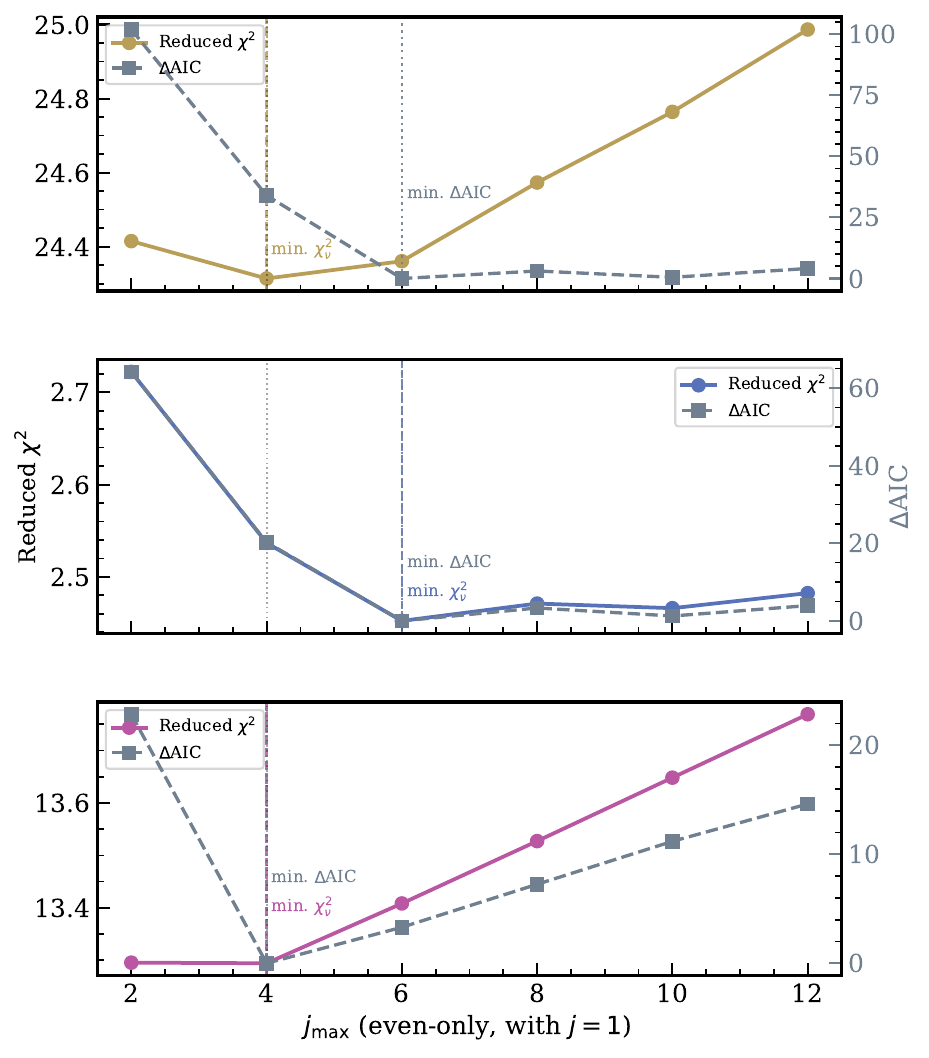}
  \caption{Model selection for longitudinal harmonic content.
  For each endmember (rows), we show reduced $\chi^2$ (solid, left axis) and $\Delta$AIC (dashed, right axis) versus $j_{\max}$ for even-only expansions (always including $j{=}1$).
  Reduced $\chi^2$ attains shallow minima at $j_{\max}{=}4$ for Endmembers~1 and~3 and at $j_{\max}{=}6$ for Endmember~2 (coloured vertical dashes), whereas AIC favours $j_{\max}{=}6$ for Endmembers~1 and ~2 and $j_{\max}{=}4$ for Endmember~3 (black dotted line). We adapt a conservative $j_{\max}{=}4$ for all endmembers, illustrated by the faint grey vertical line. 
  }
  \label{fig:appendix_jmax_criteria}
\end{figure}

\begin{figure*}[h]
  \centering
  \includegraphics[width=0.85\textwidth]{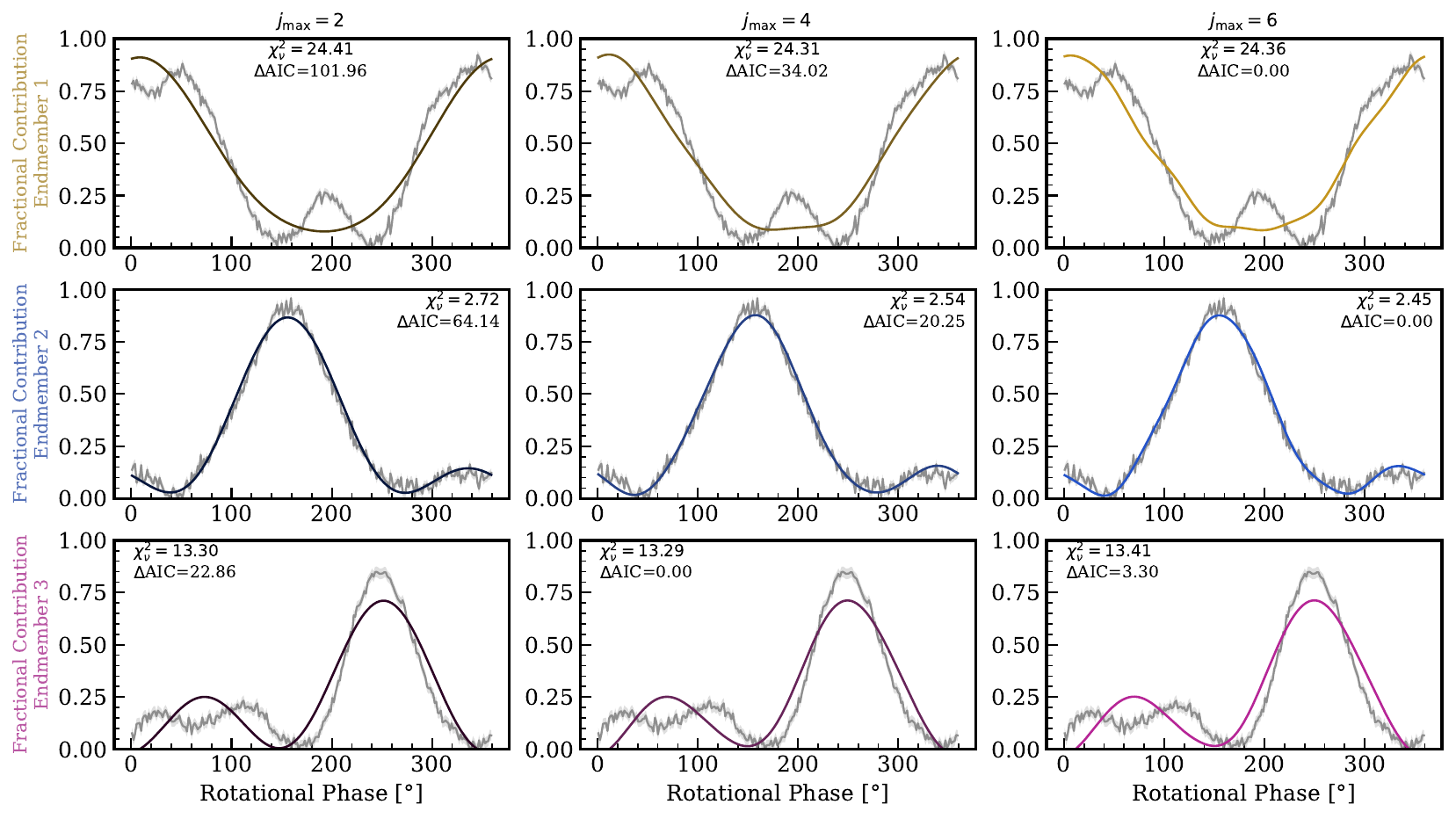}
  \caption{Fits to endmember contribution curves for increasing harmonic content.
  Rows show the three endmembers; columns show even-only models with $j_{\max}=2,4,6$. Grey lines/bands: observed contribution curves with $1\sigma$ uncertainties. Coloured lines: best-fit models. Reported $\chi^2_\nu$ and $\Delta$AIC illustrate that moving from $j_{\max}{=}2$ to $j_{\max}{=}4$ yields small but measurable improvements, while $j_{\max}{=}6$ adds little visually and is disfavoured by AIC. The close similarity of the curves across columns highlights that the first few harmonics capture most constrained structure.}
  \label{fig:appendix_jmax_fit}
\end{figure*}

\begin{figure*}[h]
  \centering
  \includegraphics[width=0.85\textwidth]{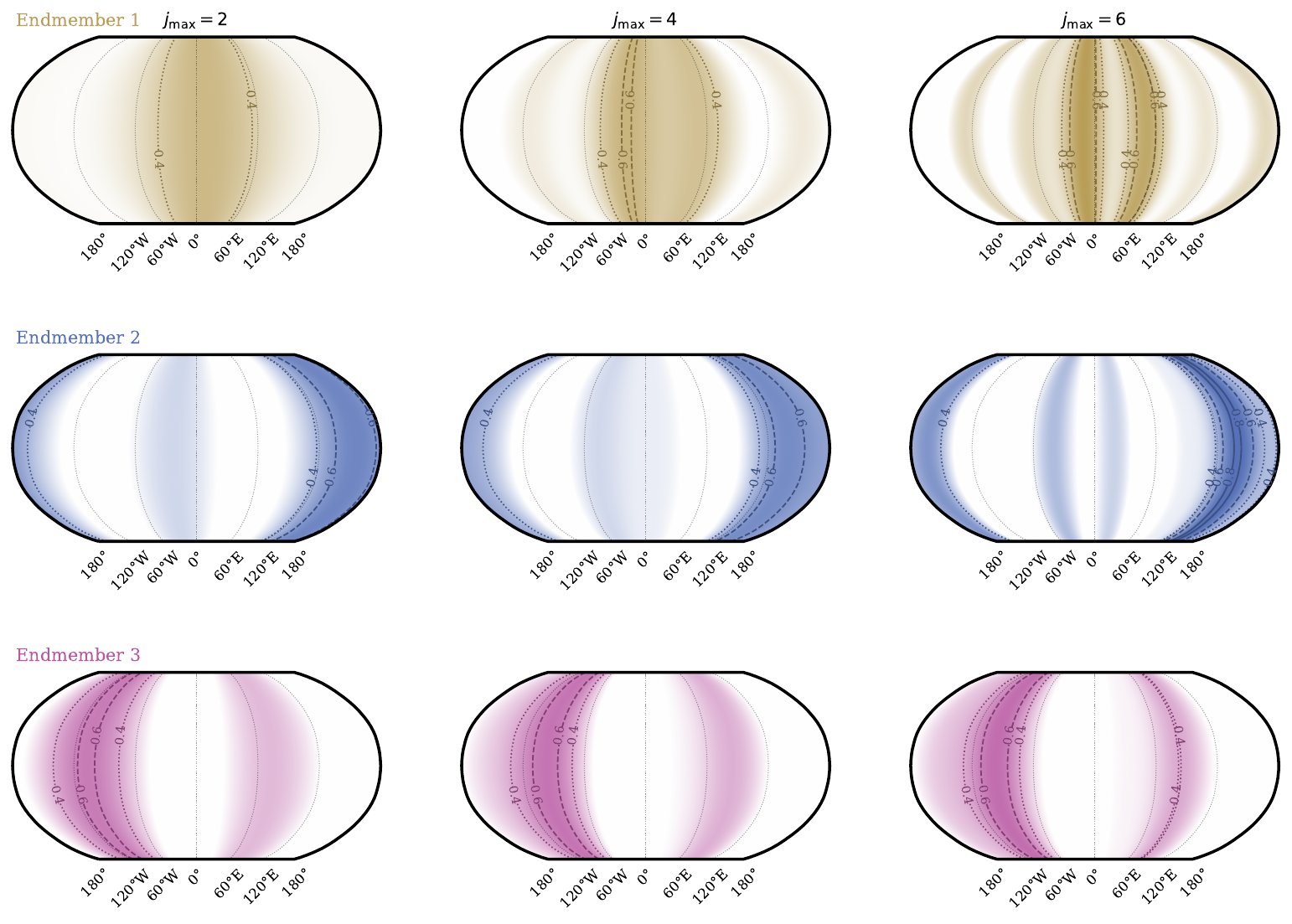}
  \caption{Robinson-projected longitudinal surface maps from Fourier inversions.
  For each endmember, we show the reconstructed brightness as a function of longitude using the adopted even-only truncation. Contours indicate fractional levels of the map. The maps are dominated by the first two harmonics, consistent with kernel smoothing that attenuates high-$j$ power; increasing $j_{\max}$ beyond 4 produces negligible morphological changes (cf. Fig.~\ref{fig:appendix_jmax_fit}).}
  \label{fig:appendix_jmax_criteria}
\end{figure*}

\end{appendix}
\end{document}